\documentclass[journal]{IEEEtran}
\usepackage{graphicx}
\usepackage{epsfig}
\usepackage{amssymb}
\usepackage{amsthm}
\usepackage{multirow}
\usepackage[table]{xcolor}
\usepackage{float}
\usepackage{caption}
\usepackage{subcaption}
\usepackage{url}
\begin{document}
\title{Sub-vector Extraction and Cascade Post-Processing for Speaker Verification Using MLLR Super-vectors}
\author{A. K. Sarkar,
        C. Barras,~\IEEEmembership{Member,~IEEE,}
        V. B. Le,
        and~D. Matrouf,~\IEEEmembership{Member,~IEEE}% <-this % stops a space
\thanks{A. K. Sarkar and C. Barras are with LIMSI-CNRS, Universit\'{e} Paris-Sud, BP 133, 91403 Orsay, France
\protect \\ V. B. Le is with Vocapia Research, 28 Rue Jean Rostand, Parc Orsay Universit\'{e}, 91400 Orsay, France \protect \\D. Matrouf is with Universit\'{e} d'Avignon, LIA, Avignon, France\protect \\
E-mail:\{sarkar,barras\}@limsi.fr, levb@vocapia.com, driss.matrouf@univ-avignon.fr}}% <-this % stops a space
\maketitle

\begin{abstract}
  In this paper, we propose  a speaker-verification system based on maximum likelihood linear regression (MLLR) super-vectors, for which speakers are characterized by m-vectors. 
These vectors are obtained by a uniform segmentation of the speaker MLLR super-vector using an overlapped sliding window. 
We consider three approaches for MLLR transformation, based on the conventional $1$-best automatic transcription, on the lattice word transcription, or on a simple global universal background model (UBM). 
Session variability compensation is performed in a post-processing module with probabilistic linear discriminant analysis (PLDA) or the eigen factor radial (EFR).
Alternatively, we propose a cascade post-processing for the MLLR super-vector based speaker-verification system.
 In this case, the m-vectors or MLLR super-vectors are first projected onto a lower-dimensional vector space generated by linear discriminant analysis (LDA). 
Next, PLDA session variability compensation and scoring is applied to the reduced-dimensional vectors. 
This approach combines the advantages of both techniques and makes the estimation of PLDA parameters easier. 
Experimental results on telephone conversations of the NIST 2008 and 2010 speaker recognition evaluation (SRE) indicate that the proposed m-vector system performs significantly better than the conventional system based on the full MLLR super-vectors. 
Cascade post-processing further reduces the error rate in all cases. 
Finally, we present the results of fusion with a standard i-vector system in the feature, as well as in the score domain, demonstrating that the m-vector system is both competitive and complementary with it.
\end{abstract}
\begin{IEEEkeywords}
m-Vector, Lattice/$1$-best MLLR, MLLR super-vector, PLDA, Speaker Verification
\end{IEEEkeywords}

\IEEEpeerreviewmaketitle

\section{Introduction}

Most state-of-the-art text-independent speaker-verification systems currently rely on the i-vector approach, where a universal background model (UBM) gaussian mixture model (GMM) representing a generic model of speakers is adapted to each target speaker by the maximum a posteriori (MAP) adaptation~\cite{reynold00}; a super-vector of the model means is further projected to a lower-dimension space, resulting in a compact i-vector representative of the target speaker~\cite{Deka_ieee2011}.
Alternatively, maximum likelihood linear regression (MLLR) super-vectors were introduced for speaker verification in a support vector machine (SVM) framework by Stolcke et al.~\cite{Stolcke05}, followed by several variants~\cite{KaramIcassp2008}. This was found to be both competitive and complementary with  the approach  of gaussian means super-vector associated with an SVM classifier~\cite{ferras_ieee2010}.
%An Automatic Speech Recognition (ASR) front-end is generally used for estimating several MLLR transformations for a given speaker speech segment with respect to pre-defined phonetic classes; MLLR transformations are then concatenated to form a MLLR super-vector. Alternatively, a single MLLR transformation can be estimated from a simple UBM, even if this leads to lower performances.
More recently, only a few studies have further explored MLLR super-vectors for speaker verification~\cite{Nicolas_Scheffer2011,Cumani2012,Stolcke2012}. %achintya-interspeech2011,

% and efficient ways of integrating MLLR super-vectors as a relevant source of information for speaker recognition.

Our aim in this paper is to explore a new representation of the speakers by their MLLR super-vectors
We propose an MLLR-based speaker-verification system, where speakers are characterized by vectors called m-vectors by analogy with i-vectors, following and extending our preliminary work~\cite{achintya-eusipco2012,achintya-icassp2013}.
These vectors are obtained by a uniform segmentation of the speaker MLLR super-vector using an overlapped sliding window.
The smaller dimension of the m-vectors compared to the entire MLLR super-vector limits the  sparsity of the data and makes session-variability compensation easier.
The experiments in~\cite{achintya-eusipco2012} were performed with m-vectors extracted from a UBM-based MLLR transformation, i.e., using a single, global model, not considering the phonetic information. 
In this work, we consider both a UBM-based MLLR transformation as well as phonetic class-based MLLR transformations; the latter is estimated  either from the conventional $1$-best automatic speech recognition (ASR) transcription or from the lattice word transcription. The lattice is indeed able to account for the ASR transcription errors, resulting in a more robust estimation of the MLLR transformations.

Second, we propose a two-stage post-processing method for the MLLR super-vector-based speaker-verification system, similar to the i-vector framework.
In this case, the MLLR-based super-vectors are first projected onto a lower-dimensional discriminant or dominant vector space, which is generated by linear discriminant analysis (LDA) or principle component analysis (PCA).
Then, probabilistic linear discriminant analysis (PLDA) session-variability compensation and scoring is applied to the reduced-dimensional vectors.
This combines the advantages of both techniques and also helps the estimation of the PLDA parameters by reducing the dimension of the representation space.

Finally, we present the fusion of the proposed m-vector technique with a standard i-vector system in the feature as well as  the score domain, showing the complementarity of the approaches.
The experimental results are presented for a standard task of the NIST 2008 and 2010 speaker recognition evaluation (SRE) core condition with English telephone conversations~\cite{nisteval08,nisteval10}.

The paper is organized as follows:
Section~\ref{sec:mllr} presents a description of MLLR and its application to speaker recognition in the m-vector framework.
We describe the reference systems and the post-processing common to all systems
in Sections~\ref{sec:ref_systems}.
Section~\ref{sec:prop_method} focuses on our proposed m-vector based systems.
The experimental setup is described in Section~\ref{sec:setup}.
Section~\ref{sec:results} reports the experimental results and discusses the performances.
Finally, Section~\ref{sec:conclusion} summarizes the work and draws several conclusions.

\section{MLLR-Based Speaker Modeling}
\label{sec:mllr}
MLLR is a speaker-adaptation technique that is mostly used in the ASR system to obtain the adapted speaker models for a given speech dataset.
The speech signal is first decoded using a speaker-independent model, and then an MLLR transformation is estimated using either the $1$-best or the lattice phonetic transcription.
However, in speaker recognition, the MLLR transformation parameters are generally used in the form of a super-vector instead of forming  a speaker-adapted model.
In this section, we describe the estimation of the MLLR transformations, the formation of the MLLR super-vector and, finally, their transformation into m-vectors for speaker verification.

\subsection{MLLR transformation}
MLLR~\cite{Leggeter95} is commonly used for the speaker adaptation of  speaker-independent (SI) hidden markov model (HMM)-based ASR systems, by estimating the affine transformations  expressed as
\begin{equation}
\hat{\mu}_s=A\mu_s + b; \quad \hat{\Sigma}_s=\Sigma_s 
\end{equation}
where ($\mu_s$, $\Sigma_s$) and  $(\hat{\mu}_s, \hat{\Sigma}_s)$ represent the gaussian mean and covariance matrix of the $s^{th}$ state in the SI and adapted model, respectively.
$(A,b)$ is called the \emph{MLLR transformation}.
%In our experiment, bias $b$ is not used.
%Following steps are involved in the estimation of $A$:
%\begin{itemize}
%\item[]{\bf Step 1:} Calculate the occupation likelihood, ${\gamma}_{j}\left(t\right)$  for the  $j^{th}$ Gaussian  for given feature vectors $X_r$ at time $t$
%\item[]{\bf Step 2:} Compute the following two sufficient statistics for \emph{$i^{th}$ components (dimension) of feature vectors},
%         \begin{eqnarray}
%          {K^{(i)}} & = & \sum_{j=1}^{c}\sum_{t=1}^{T} \gamma_{j}(t) \; \frac{1}{\sigma_{ji}^{2}} \;  x_{i}(t) \; \mu_{j}^{'} \label{eq:EM-SS-K} \\
%         {G^{(i)}} & = & \sum_{j=1}^{c} \frac{1}{\sigma_{ji}^{2}} {\mu_{j}}\mu_{j}^{'} \sum_{t=1}^{T}{\gamma}_{j}\left(t\right) \label{eq:EM-SS-G}
%       \end{eqnarray}
%$\mu_{j}$ and ${\sigma_{ji}^2}$ denote the mean and the $i^{th}$ component of the covariance matrix for the  $j^{th}$ Gaussian, respectively. The symbol $(.)'$ indicates matrix transpose operation.
%\item[]{\bf Step 3:}  \emph{$i^{th}$ row of the $A$ transformation} is obtained as,
%       \begin{equation}
%        A_i= K^{(i)} {G^{(i)}}^{-1} \label{eq:MLLR_tran} 
%       \end{equation}
%\item[]{\bf Step 4:} Repeat Step 2 to 3 upto feature vector dimension \\ %(i.e. $47$ in our experiment) 
%\end{itemize}
%
MLLR transforms are usually estimated across a set of gaussians that share identical transformation parameters. In the context of an ASR system, these classes can be defined thanks to the phonetic similarities of the acoustic models and may represent phonetic classes.
Therefore, each of the regression classes results in a separate MLLR transform.

%\subsection{Lattice MLLR}
Furthermore, automatic transcriptions of telephone conversations present typical word-error rates (WER) in the range of $20$--$30\%$. 
Therefore, MLLR transformation estimated based on the $1$-best hypothesis often misses the correct acoustic model.
To account for the transcription errors, lattice-based MLLR transforms~\cite{padmanabhan_00,uebel_01} are estimated using the word-lattice output of an ASR system obtained by first-pass decoding, which is converted into a model-level graph  using the pronunciation variants in the lexicon.
%It can be expressed as,
%\begin{equation}
%\hat{\theta} =\mbox{arg }\max_{\theta=(\mathbb{A}, \mathbb{B}, \pi )} \sum_{s_r\epsilon {\bf S}} p(s_r|X_r,\theta)\; log\; p(X_r, s_r |\theta) \label{eq:lattice_mllr}
%\end{equation}
%where $p(s_r|X_r,\theta)$ represents the probability of aligning the $r^{th}$ speaker training data, $X_r$ with respect to state sequence $s_r$ using SI HMM model, $\theta$.
%{\bf S} indicates all possible alignment state sequences of $X_r$ with respect to the SI model.
%In the conventional approach in Eq.~\ref{eq:ibest_mllr}, $p(s_r|X_r,\theta)$ is set to $1$ for the $1$-best hypothesis (i.e. $1$-best state sequence) and $0$ for others.
Details about the use of the lattice MLLR approach for speaker verification can be found in~\cite{marc2009}.

%\subsection{MLLR super-vector}
%As illustrated by Fig. ~\ref{fig:mllr_sup}, 

\subsection{m-vectors extraction}
\label{sec:m-vectorTech}

An MLLR super-vector~\cite{Stolcke05} is formed by stacking the elements of the MLLR transformation matrix $A$, e.g., row-wise.
%\begin{figure}[H]
%\centering{\includegraphics[height=3.5cm,width=7.5cm]{figs/mllr_super_vector.eps}}
%\caption{\it MLLR super-vector estimation from a speaker independent HMM using the training data of the $r^{th}$ speaker.}
%\label{fig:mllr_sup}
%\end{figure}
%
The bias $b$ did not provide any significant gain in our experiments and is not considered further.

\begin{figure}[h]
\centering{\includegraphics[height=3.8cm,width=7.5cm]{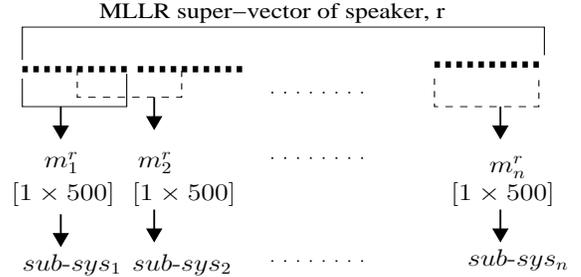}}
\caption{\it \emph{m-vector} extraction for the $r^{th}$ speaker from his/her MLLR super-vector using an overlapped sliding window of $500$ elements with $50\%$ overlap of its adjacent \emph{m-vectors}.}
\label{fig:m-vector}
\end{figure}

The m-vector technique has been recently proposed for speaker verification~\cite{achintya-icassp2013}.
In this approach, the speakers are characterized by a set of m-vectors, which are extracted from their MLLR super-vectors by a uniform segmentation using overlapping, sliding windows, as shown in Fig.~\ref{fig:m-vector}.
The following assumptions  motivate  the m-vector technique.
Each row of the MLLR transformation is associated with a particular dimension of the feature vectors. Hence, each m-vector will capture the speaker-relevant information related to a subset of the features in a more compact way than the full MLLR super-vector. 
Furthermore, the overlap between the adjacent m-vectors limits the impact of the segmentation process.
In contrast to the full super-vector, m-vectors have a smaller dimension, and hence, the parameter estimation of the post-processing step is less likely to be affected by the data sparsity.

A speaker is represented by several m-vectors that are processed separately and thus constitute several sub-systems, as illustrated on Fig. ~\ref{fig:m-vector}. The window size and the overlap control the actual number of m-vectors extracted from an MLLR super-vector.
When the size of the MLLR super-vector is not a multiple of the window size, an additional \emph{m-vector} is extracted by placing a window at the end point of the super-vector, to cover all elements of the super-vector.

%During test, m-vectors of the test utterance are generated in a similar manner and scored against the claimant specific m-vectors obtained during the training phase.
%Before scoring, m-vectors are post-processed for session variability compensation.

\section{Reference Systems and Post-Processing}
\label{sec:ref_systems}
In this section, we describe a set of baseline MLLR systems and a state-of-the-art i-vector system used for comparisons, and we consider various post-processing techniques for dimension reduction and session-variability compensation.

\subsection{i-vector system}
\label{sec:i-vector}
The i-vector system is based on the state-of-the-art technique for speaker-verification \cite{Deka_ieee2011}, in which speakers or speech segments are characterized by a vector called an i-vector.
i-vectors are estimated by projecting the speaker data onto a \emph{total variability space}, $T_{total}$.
This is generally expressed as,:
\begin{eqnarray}
\hat{\mu}_{gsv}=\mu_{gsv} +T_{total}w
\end{eqnarray}
where $w$ denotes an i-vector. $\hat{\mu}_{gsv}$ and $\mu_{gsv}$ are the GMM super-vectors of the speaker-dependent and speaker-independent (i.e., UBM) model, respectively. 
$T_{total}$ is a low-rank matrix representing the sub-space of the GMM super-vector domain, assuming that the speaker and channel variabilities are concentrated.
%$c$, $F$ and $R$, are respectively the number of components in the UBM, the dimension of the feature vector and of the i-vector.

The i-vector framework greatly simplifies the speaker-recognition problem, because the similarity between the utterances from two different speakers may be simply computed using a Mahalanobis distance, instead of computing a log-likelihood ratio between the models given the data. The simplicity and efficiency of this approach led to applications into various domains, e.g., for the representation of HMM states for clustering%~\cite{Bouallegue2012}
or the classification of segments for speaker segmentation~\cite{Gregor2012}. This approach was also used for modeling the prosodic information~\cite{Kockmann2012} and the phonotactic information~\cite{Soufifar2013} for language recognition. The impact of the training segment duration for the i-vectors was explored in~\cite{Larcher-icassp2012},%achintya-interspeech2012,
and a method for balancing this effect was proposed in~\cite{Taufiq-icassp2013}.

\subsection{Baseline MLLR-based systems}
\label{sec:baseline}
We consider several baseline systems based on the full MLLR super-vector, for a fair comparison with the proposed methods derived from the same super-vector. 
Four configurations for the dimension reduction of the MLLR super-vectors are compared: 
\begin{itemize}
\item \emph{LDA}, \emph{PCA} and \emph{PPCA-NAP}, in which speakers are characterized by projecting their \emph{full} MLLR super-vectors in a lower-dimensional space using LDA, PCA or probabilistic PCA with nuisance attribute projection (PPCA-NAP), followed by eigen factor radial (EFR) session-variability compensation and scoring as described below.
\item \emph{PLDA}, where the full MLLR super-vectors are scored in the PLDA space.
\end{itemize}
In each case, depending on the models and procedure used for the MLLR estimation, three types of systems result, namely \emph{full ASR $1$-best system}, \emph{the full ASR lattice system} and \emph{the full UBM system}.

\subsection{Eigen factor radial (EFR)}
\label{sec:efr}
EFR is a session-variability compensation (i.e., post-processing) and scoring technique. It was introduced in \cite{Pierre-interspeech2011} to handle the session-variability compensation by iterative length normalization of the i-vector (i.e., $w$) as:
 \begin{equation}
 \hat{w} \leftarrow \frac{V^{-\frac{1}{2}}(w-\overline{w})}{\sqrt{(w-\overline{w})^{t} V^{-1}(w-\overline{w})}} \label{eq:cond_}
 \end{equation}
where $V$ and $\overline{w}$ denote the covariance matrix and mean vector of the training i-vectors,  respectively, in successive iterations.
$V$ and $\overline{w}$ are estimated from data collected over many non-target speakers.
 
During a test, the score between two post-processed i-vectors (e.g., $\hat{w}_1, \hat{w}_2$) is calculated using the Mahalanobis distance:
\begin{equation}
score(\hat{w}_1,\hat{w}_2) = (\hat{w}_1-\hat{w}_2)^{t} \Omega^{-1}(\hat{w}_1-\hat{w}_2)
\label{eq_maha}
\end{equation}
where $\Omega$ denotes the within-class covariance matrix calculated over the non-target speakers data set.

EFR was shown to give better performance than the conventional LDA followed by within class covariance normalization (WCCN) with a cosine angle for speaker verification~\cite{Pierre-interspeech2011}.
%
% In our case, m-vectors/full MLLR super-vectors are first projected on the LDA space to reduce the dimension and discriminant among the speakers.
% Then, EFR is applied on them to handle the session variability compensation.
%
Thus, EFR is applied to both the m-vector and the MLLR super-vector systems for session-variability compensation and scoring.
In the m-vector case, each sub-system has its own $V$, $\overline{w}$ and $\Omega$.
Two iterations of length normalization  (i.e., Eq. \ref{eq:cond_}) are considered for all systems presented in the paper using EFR.

\subsection{Probabilistic linear discriminant analysis (PLDA)}
PLDA is also a session-variability compensation and scoring technique. It is basically
a generative modeling technique that decomposes the i-vector (i.e., $w$) into several contributions: %like Joint Factor Analysis (JFA) framework.
\begin{eqnarray}
w=\mu_w +\phi \;y_s + \Gamma \;z + \epsilon
\end{eqnarray}
where $\phi$ and $\Gamma$ represent the eigen-voice and eigen-channel subspaces, respectively.
$y_s$, $z$ and $\epsilon$ are the speaker factor, channel factor and residual noise, respectively.

During a test, the score between two i-vectors (e.g., $w_1$, $w_2$) is calculated as:
\begin{eqnarray}
score(w_1,w_2) = log\;\frac{p(w_1,w_2|\theta_{tar})}{p(w_1,w_2|\theta_{non})}
\end{eqnarray}
where the hypothesis $\theta_{tar}$ defines that $w_1$ and $w_2$ are from the same speaker,
and $\theta_{non}$ states that $w_1$ and $w_2$ are from different speakers.
In our case, PLDA is also applied on the m-vectors or MLLR super-vectors.
For details about the training of the PLDA parameters ($\phi,\epsilon, \theta_{tar}, \theta_{non}$), see \cite{Prince2012,SenoussaouiInterspch2011}.
As per \cite{Pierre-odyssey2012}, two iterations of length normalization following Eq.~\ref{eq:cond_} are applied to the data before PLDA. 

\subsection{PPCA-NAP}
The PPCA-NAP approach consists of removing an eigen-channel subspace from the full MLLR super-vectors for speaker characterization, similar to eigen-channel compensation on the MLLR super-vectors \cite{Nicolas_Scheffer2011}. 
MLLR super-vector $o$ is decomposed as:
%like eigen channel algorithm for cepstral features as,
  \begin{eqnarray}
   o&=&o_s + Uy \\
   y&=&P^{-1}U^{'}o ; \quad P=I+U^{'}U
  \end{eqnarray}
where $y$ is the point estimator and $U$ is the low-rank intra-speaker variability matrix having prior distribution ${\cal N}(0,1)$.
The $U$ matrix is estimated with $30$ iterations of maximum likelihood (ML) estimation. 
%  
%During test, after removal of the eigen channel subspace, the super-vector of the test utterance is scored against the claimant.
We use EFR-based scoring, which yields better performance in our experiments than the inner product chosen by~\cite{Nicolas_Scheffer2011}.
%  It is also noted that it is already shown in \cite{Pierre-interspeech2011} that EFR gives better speaker verification performance than LDA+WCCN and cosine scoring.
          
\subsection{Test phase}
During the test phase, the m-vector, i-vector or MLLR super-vector of the test utterance is scored against the claimant specific vector, after post-processing of both vectors.
% ??? not clear to me
%In the m-vector case and early fusion, the scores of the subsystems are fused %across the particular LDA/PLDA(speaker and channel factors).
For the m-vector systems combining several subsystems, the scores of the subsystems are equally weighted for fusion, i.e.,
\begin{equation}
score_{fusion} =  \frac{1}{N_{subsys}} \sum_{i=1}^{N_{subsys}} score(\tilde{m}_i^{r},\tilde{m}_i^{test}) 
\end{equation}
where $\tilde{m}_i^{r}$ and $\tilde{m}_i^{test}$ represent the post-processed \emph{m-vectors} of the claimant, $r$ and test utterance for the $i^{th}$ subsystem, respectively.
$score(.,.)$  denotes the scoring function between the two \emph{m-vectors}.

\section{Proposed Approaches}
 \label{sec:prop_method}
We propose in this section three variants of m-vector systems, along with a cascade post-processing for reducing the dimension of the MLLR super-vectors.
  
\subsection{m-vector systems}
 \label{sec:1best_mllr}
  For the \emph{ASR $1$-best m-vector system} and the \emph{ASR lattice m-vector system}, MLLR transformations are estimated with respect to an SI HMM using the conventional $1$-best hypothesis or the lattice transcription for a given speech segment, respectively.
  Then, class-wise MLLR transformations grouped into a super-vector are used for speaker characterization by m-vectors during a training session,
 as described in Section~\ref{sec:m-vectorTech}.
In this work, $42$ dimensional feature vectors are used and MLLR transformations are estimated with respect to two predefined phonetic classes, vowel and consonants.
This results in a $42 \times 42=1764$ -dimensional MLLR super-vector for each phonetic class and finally a $2 \times 1764=3528$ dimensional MLLR super-vector for a given speech segment.
  During the test, the m-vectors of the test utterance are scored against the claimant. 
Before scoring, the m-vectors are post-processed for session-variability compensation and scoring with PLDA or EFR.
In the case of EFR, LDA is first applied on the m-vectors to reduce their dimension and improve the discrimination between the speakers.
LDA is implemented independently for each sub-system.
Hence, each sub-system has its own LDA projection matrix.

 The \emph{UBM m-vector system} is similar to the \emph{ASR $1$-best or lattice m-vector systems}.
The main difference is that a simple UBM is considered as the SI model.
 A single-class, global MLLR transformation is estimated with respect to the UBM for a given speaker dataset \emph{without} any speech transcription or phonetic knowledge.
This results in a $42 \times 42=1764$ -dimensional MLLR super-vector.
 
\subsection{Cascade post-processing}
\label{sec:cascade}
\begin{figure}[h]
\centering{\includegraphics[height=3.1cm,width=8.8cm]{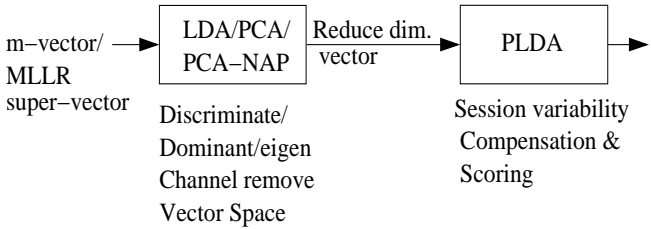}}
\caption{\it Proposed cascade post-processing for speaker verification using m-vectors or full MLLR super-vectors.}
\label{fig:cascade_system}
\end{figure}

PLDA is commonly used in state-of-the-art speaker-verification systems using an i-vector for the  session-variability and scoring technique without applying prior PCA, or LDA or PPCA-NAP to the data.
In contrast to that of the i-vector, the dimension of the MLLR super-vectors is larger, and a direct estimation of the PLDA parameters may raise estimation issues due to the limited training examples.
To reduce this risk, we propose a cascade post-processing, in which the m-vectors or full MLLR super-vectors are first projected onto a discriminant or dominant lower-dimensional vector space generated by LDA or PCA, respectively.
Then, PLDA is applied on these reduced-dimensional vectors for session-variability compensation and scoring, as shown in Fig. ~\ref{fig:cascade_system}.

%The main difference of the baseline systems proposed in Section~\ref{sec:baseline} with this \emph{cascade system} is that no further LDA or PLDA is applied on top of the MLLR projection.
 
\section{Experimental Setup}
\label{sec:setup}
All experiments are performed on male speakers of the seventh common condition of interest (all trials involve only English language telephone speech in the training and the test) of NIST SRE 2008 core condition and of the fifth task (telephone-only) of the SRE 2010 core condition~\cite{nisteval08,nisteval10}.
There are $1270$ and $5200$ target models in NIST SRE 2008 and 2010, respectively. Each target is provided a single utterance for training its model. The utterances are approximately 5 minutes with $2.5$ minutes of speech on average. 

For signal analysis, $42$-dimensional vectors, including the $12$ Mel-PLP feature, log-energy and $F_0$ along with their first- and second-order derivatives, are extracted from the speech signal 
at a $10$-ms frame rate using a $30$-second Hamming window over the $0$-$3800$~Hz bandwidth.
Voice activity detection is applied as a pre-processing step to discard less energetic or silent frames.
Finally, the selected frames are normalized to a zero mean and unit variance at the utterance level.
Two manually derived phonetic classes, vowels and consonants, are used for the MLLR transformations, estimated on the PLP+F$_0$ features only. Audio segments aligned with the silence model after the decoding are not considered for the MLLR transformation relying on the ASR transcriptions, but are retained for the UBM-based system.
All MLLR transformations are estimated with a single iteration.

A male-specific gender-dependent UBM with 512 gaussians and diagonal covariance matrices is trained using data from NIST SRE 2004.
The large vocabulary continuous speech recognition (LVCSR) system used for MLLR transforms estimation is similar to the LIMSI RT'04 LVCSR system~\cite{Prasad05}. 
The acoustic models are trained using approximately 2000 hours of manually transcribed conversational telephone speech (CTS) data using the PLP+F$_0$ features concatenated with additional MLP features~\cite{Fousek2008}.
The model sets cover approximately $48k$ phone contexts, with $11.5k$ tied states and $32$ Gaussians per state. 
Silence is modeled by a single state with $1024$ Gaussians.

The LDA, PLDA, PCA, PPCA-NAP and EFR algorithm are implemented using data from $890$ non-target speakers from NIST $2004$-$2005$, Switchboard II parts $1$, $2$ \& $3$, Switchboard cellular parts $1$ \& $2$, with approximately $15$ sessions per speaker.
This results in $12392$ utterances i.e., $12392$ MLLR super-vectors.
This data-set is also used for the training total variability space of the $400$-dimensional i-vector system.
In the cases of PCA, PPCA-NAP and LDA, the m-vectors or full MLLR super-vectors are normalized to a zero mean and unit variance. % at each vector level before using them, which gives the best performance for the respective systems.
For PLDA, both the speaker- and channel- factor dimensions are varied from the dimension of the initial vector (m-vector, i-vector or full MLLR super-vector) with a step of $50$  to determine the optimal performance of the systems.
%This data set is also used for training total variability space in i-vector system.
%
For the SRE 2010 experiments, $6947$ additional utterances are taken from  NIST SRE 2006 and 2008 for T-space, LDA and PLDA implementation.

All system performances are evaluated in terms of the equal error rate (EER) and minimum detection cost function (MinDCF), following SRE 2008 and 2010 evaluation~\cite{nisteval08,nisteval10}.

\section{Results and Discussions}
\label{sec:results}
For analysis, the speaker-verification performance of the systems are compared in terms of EER on task 7 of NIST SRE 2008 using the EFR post-processing and scoring technique.
The comparison between the proposed method and the baseline system are presented for various post-processing techniques on NIST SRE 2008 and on the most similar task of the SRE 2010 core condition.

\subsection{Performance of the baseline systems}
\label{sec:opt_baseline_system}

\begin{figure}
%\centering
\begin{subfigure}[h!]{9.0cm}
%\centering
%\hspace*{-3.0cm}
\includegraphics[height=4.5cm,width=8.0cm]{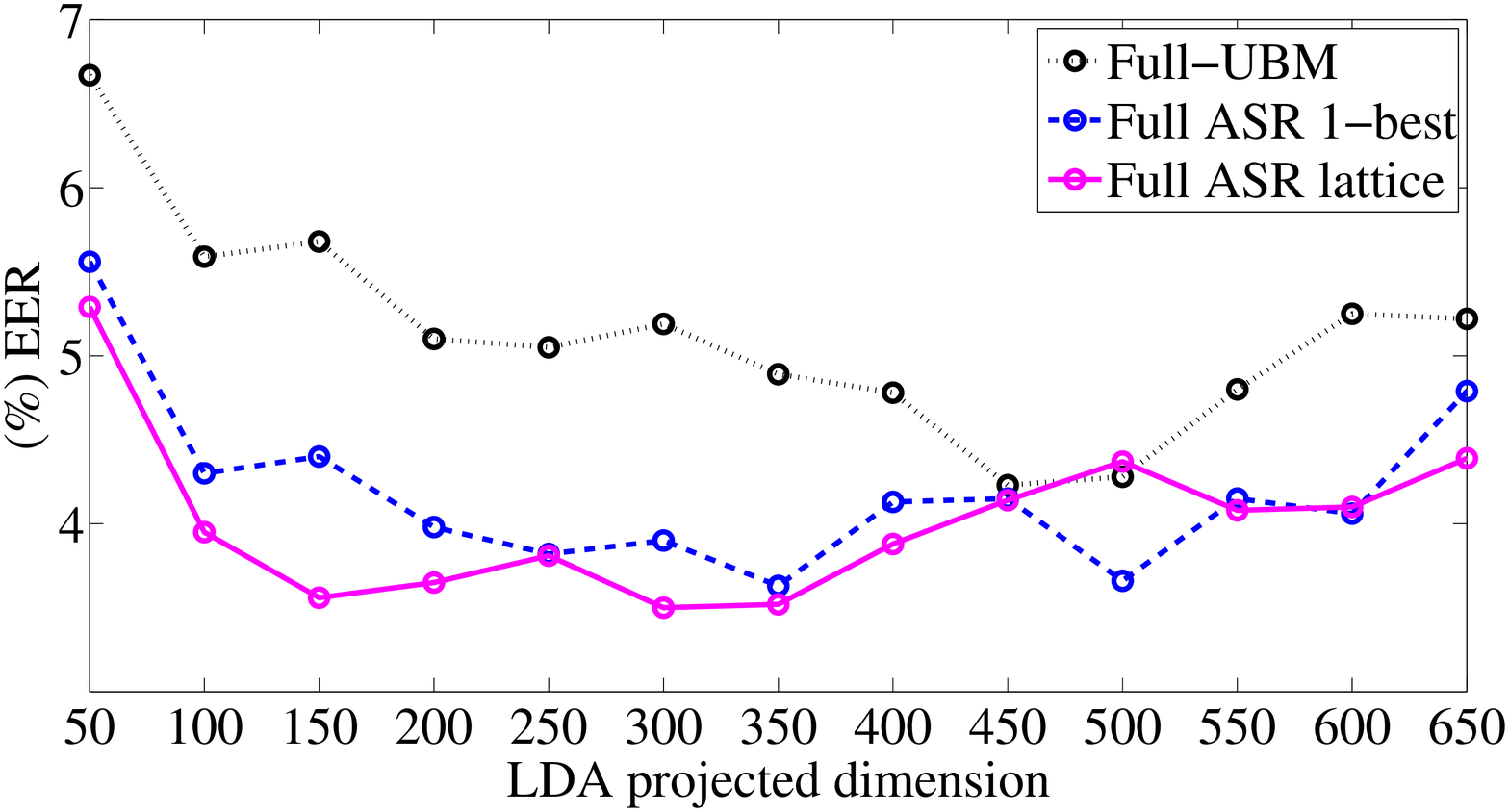}
\subcaption{\it with LDA}
\label{fig:full_LDA}
\end{subfigure}
\\[3.0ex]
\begin{subfigure}[h!]{9.0cm}
%\centering
%\hspace*{-3.0cm}
\includegraphics[height=4.5cm,width=8.0cm]{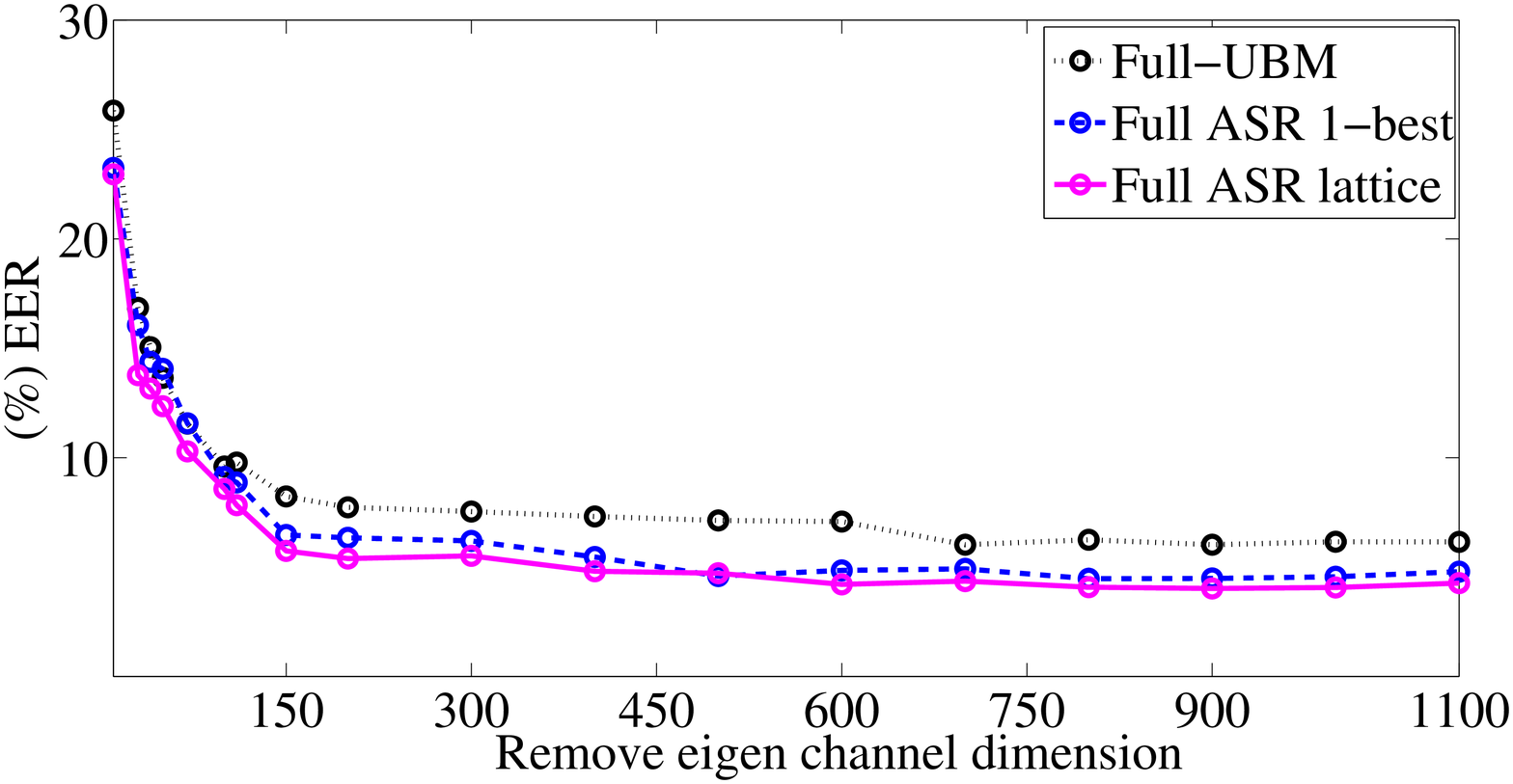}
\subcaption{\it with PPCA-NAP} 
\label{fig:full_PCA_NAP}
\end{subfigure}
\\[3.0ex]
\begin{subfigure}[h!]{9.0cm}
%\centering
%\hspace*{-3.0cm}
\includegraphics[height=4.5cm,width=8.0cm]{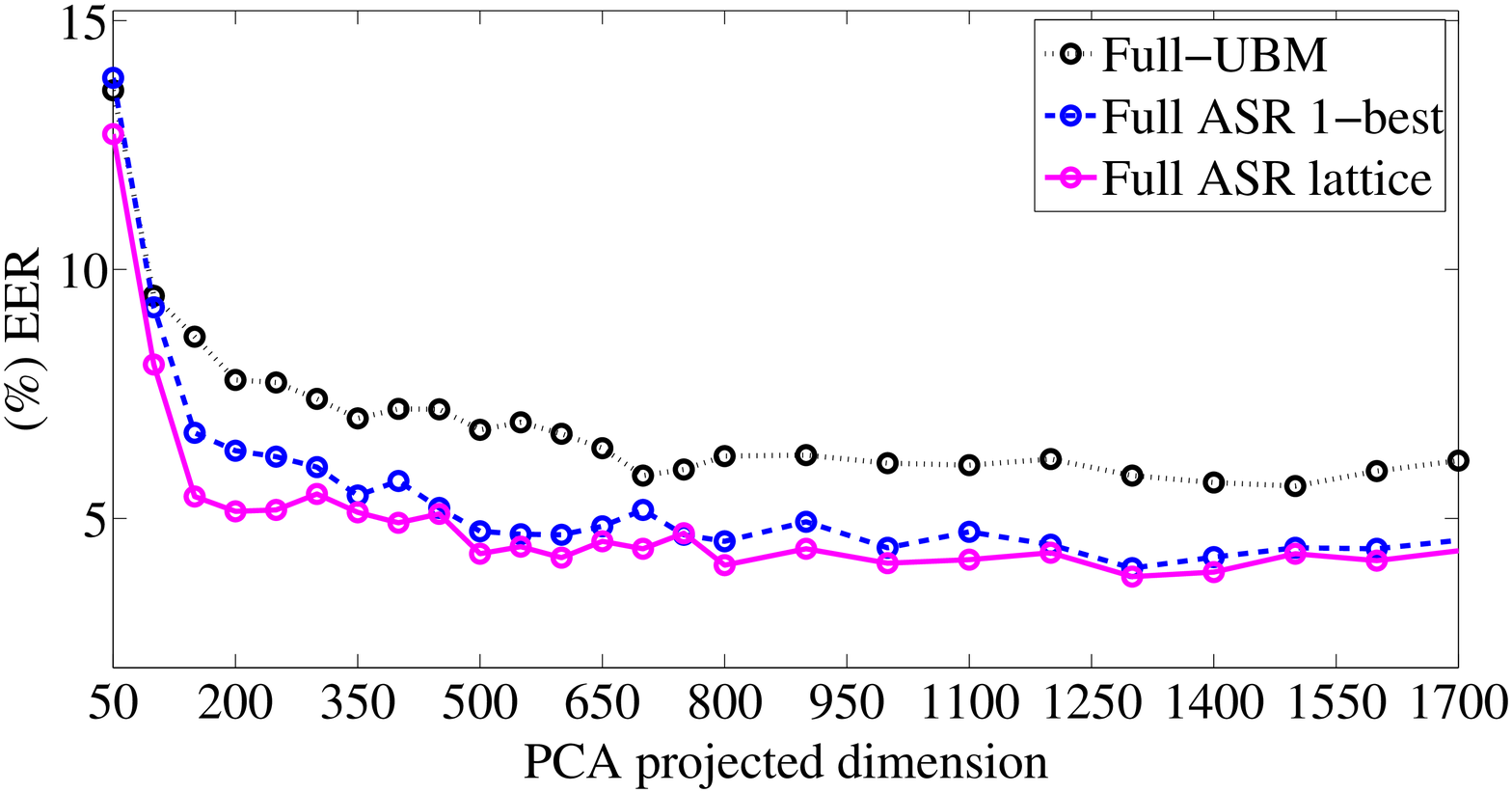}
\subcaption{\it with PCA} 
\label{fig:full_PCA}
\end{subfigure}
\caption{\it Speaker-verification performance of the baseline systems with the full MLLR super-vector for various projection dimensions using LDA, PPCA-NAP or PCA on the respective systems on task 7 of the NIST SRE 2008 core condition with EFR post-processing and scoring.}
\label{fig:ldapcanap}
\end{figure}

The speaker-verification performance of the baseline systems with PCA, PPCA-NAP and LDA is shown in Fig. \ref{fig:ldapcanap}, and an optimal projection is selected for each case according to the lowest EER.
%
%For  LDA, it can be observed from Fig.~\ref{fig:full_LDA} that the EER value of the full-UBM system decays for increasing LDA order upto the optimal LDA projection dimension $450$ for this system with 4.23\% EER and afterward again start increasing. 
%For the full ASR $1$-$best$ system, EER decreases upto the optimal LDA dimension of $350$ with 3.63\% EER.
%
%Similarly, the full ASR lattice MLLR system shows best system performance for an LDA dimension of $300$ (3.50\% EER).
%As can be seen from Fig. \ref{fig:full_PCA_NAP}, the PPCA-NAP configuration shows the lowest EER in speaker verification performance for a dimension of eigen channel removal from the MLLR super-vector of $700$, $800$ and $900$ respectively, in full UBM (6.02\% EER), full ASR $1$-$best$ (4.47\% EER) and full ASR lattice (4.02\% EER) systems.
%
%A similar pattern can be observed for PCA on Fig. \ref{fig:full_PCA}, i.e. 
%it yields the best system performance for PCA projected dimension of full MLLR super-vector to $1500$, $1300$ and $1300$ respectively, for full UBM ($5.65\%$ EER), full ASR $1$-$best$ ($4.00\%$ EER) and full ASR lattice ($3.83\%$ EER) based systems.
%
Table \ref{table:table2} summarizes the performance of the baseline systems for their respective optimal LDA, PPCA-NAP and PCA dimensions as shown in Figure~\ref{fig:ldapcanap}.
The following observations can be made:
LDA-based systems perform better than PPCA-NAP- or PCA-based systems,
whereas the performance of the systems based on PCA and PPCA-NAP is comparable.
ASR-based systems give better performance than UBM-based systems, because they incorporate the phonetic knowledge available in the speech signal onto the MLLR transformations in contrast to the UBM-based system.
As expected, ASR lattice systems perform slightly better than the conventional $1$-best ASR-based systems, with 3.5\% EER in the LDA configuration.

\begin{table}[!t]
\caption{\it Performance of the baseline MLLR super-vector systems for their respective optimal LDA, PPCA-NAP and PCA dimension on task 7 of the NIST SRE 2008 core condition with EFR post-processing and scoring.}
\centering
\begin{tabular}{l*{6}{r}r}\\ \hline
Baseline       & MLLR sup   &Proj.             & Optimal    & \%EER \\
Systems        &vec. dim.   & Method           & Proj. dim. &  \\ \hline \hline
Full UBM       & 1764          & {\bf LDA}     & 450        & {\bf 4.23} \\
               &               & PPCA-NAP       & 700        & 6.02 \\
               &               & PCA           & 1500       & 5.65  \\ \\ %\hline
Full ASR       & 3528          & {\bf LDA}     & 350        & {\bf 3.63}   \\
 1-best        &               & PPCA-NAP       & 800        & 4.47   \\
               &               & PCA           & 1300       & 4.00 \\ \\ %\hline
Full ASR       & 3528          & {\bf LDA}     & 300         &  {\bf 3.50}  \\
Lattice        &               & PPCA-NAP       & 900        & 4.02   \\
               &               & PCA           & 1300       & 3.83 \\ \hline \hline
\end{tabular}
\label{table:table2}
\end{table}

\subsection{Optimal m-vector size}
\label{sec:opt_m_vectorsize}
\begin{figure}[!t]
\centering{\includegraphics[height=4.5cm,width=8.0cm]{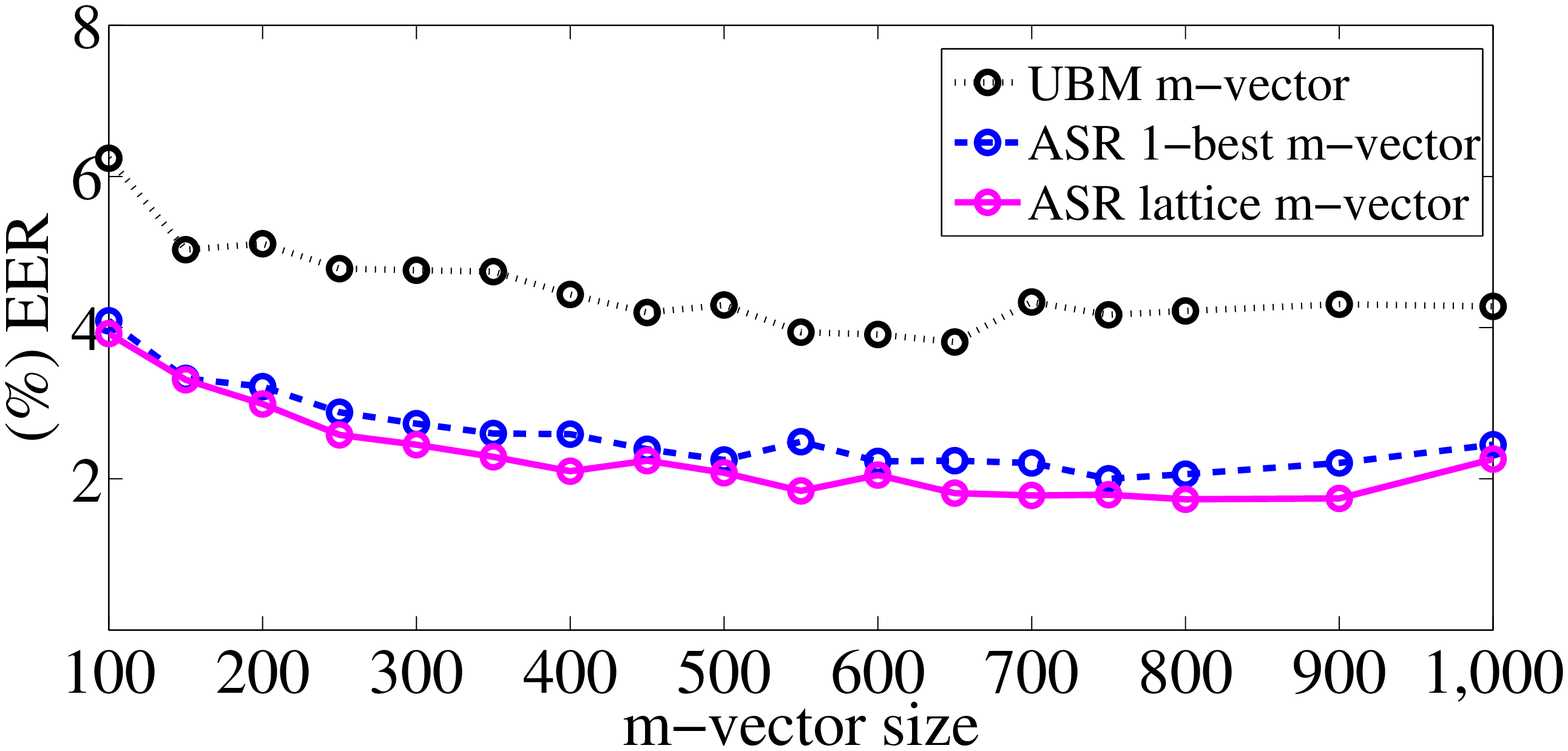}}
\caption{\it Speaker verification performance of m-vector systems in terms of EER for various m-vector dimensions on task 7 of the NIST SRE 2008 core condition, with EFR post-processing and scoring.}
\label{fig:otptm-vectordimension}
\end{figure}

We show the speaker-verification performance for a wide range of m-vector window sizes in terms of the lowest EER for the SRE 2008 common condition 7 in Fig. ~\ref{fig:otptm-vectordimension}.
For simplicity, the optimal LDA projection dimension associated with each system for a particular m-vector dimension is not shown.
The lowest EER value is achieved for  m-vector dimensions of $650$, $750$ and $800$, respectively, in the UBM, ASR $1$-best and ASR lattice m-vector cases.
Hence, these optimal m-vector dimensions are selected for the respective systems and considered afterwards in this paper.

\begin{table}[!t]
\caption{\it Performance of the m-vector systems for their respective optimal m-vector dimensions on task 7 of the NIST SRE 2008 core condition with EFR post-processing and scoring.}
\centering
\begin{tabular}{l*{6}{c}r} \\ \hline 
 Systems               &Optimal   & Optimal       & \%EER \\
                       & m-vector & LDA proj.     &  \\
                       & dim.     &  dim.         &         \\ \hline \hline
UBM m-vector           & 650      & 400           & {\bf 3.81}  \\
ASR $1$-best m-vector  & 750      & 300           & {\bf 2.00} \\
ASR lattice m-vector   & 800      & 300           & {\bf 1.73}  \\ \\ %\hline 
UBM cross m-vector     & (650+650)& 250           & 4.02  \\
ASR $1$-best cross m-vector&(750+750) & 450       & 2.02 \\
ASR lattice cross m-vector & (800+800) & 400      & 1.78 \\ \hline \hline
\end{tabular}
\label{table:table1}
\end{table}

Table \ref{table:table1} summarizes the speaker-verification performance of the m-vector systems for their optimal m-vector dimensions, as shown in Fig. ~\ref{fig:otptm-vectordimension}.
We can observe that
ASR-based m-vector systems perform significantly better than the UBM-based m-vector system, and that the ASR lattice system shows a slightly lower EER value than the $1$-best ASR m-vector system as was already observed for the baseline systems, at 1.73\% EER.

In addition, we also present the performance of the systems when the speakers are characterized by the vectors $m_{ij}$ formed by concatenation of two m-vectors $m_i, m_j$ as
%\begin{equation}
$m_{ij}=\left[m_i \; m_j \right],\; \forall i \neq j$.
%\end{equation}
The motivation of this approach is to see whether the cross-correlation among the m-vectors is able to provide a further gain in the speaker verification. 
We call this  the \emph{cross m-vector} system.
The performance of the cross and conventional m-vector systems given in Table~\ref{table:table1} is comparable, showing that the conventional approach is sufficient to capture the speaker relevant information available in the MLLR super-vector for speaker verification, at least in the framework of linear fusion and transformation.
In the remainder of the paper, only conventional m-vector systems are considered.

\begin{table*}[!t]
%\begin{center}
\centering
\caption{\it Comparison of the proposed m-vector systems with the baseline systems for standard and cascade post-processing techniques and scoring, for task 7 of the NIST SRE 2008 core condition.}
\begin{tabular}{lr|rr|rr||rr}\hline
 System  &  & \multicolumn{2}{c|}{(a) {\bf EFR} post-processing} 
 & \multicolumn{2}{c||}{(b) {\bf PLDA} post-processing} 
 & \multicolumn{2}{c}{(c) {\bf Cascade} post-processing}\\ \cline{3-8}
                           &Vector & Opt. &          & Opt. (spkr.,    &        & Opt. LDA proj.     &          \\ 
                           & dim. & LDA proj. & \%EER & chan.) factors   & \%EER  & / PLDA factors   & \%EER     \\ \hline \hline
 Full-UBM                  & 1764 & 450 & 4.23      & (1450, 1000) & 4.43       & 450 / (300,350) & {\bf 3.87} \\
{\bf UBM m-vector}         & 650  & 400 &{\bf 3.81} & (150,500)    & {\bf 3.50} & 400 / (200,350) &  3.55 \\
&&&&&&&\\
Full ASR 1-best            & 3528 & 350 & 3.63      & (850,750)    & 5.44       & 350 / (350,350) & {\bf 3.23} \\
{\bf ASR 1-best m-vector}  & 750  & 300 &{\bf 2.00} & (500,500)    & {\bf 1.91} & 300 / (300,250) & {\bf 1.80} \\
&&&&&&&\\
Full ASR lattice           & 3528 & 300 & 3.50      & (1100,700)   & 4.60       & 300 / (200,250) & {\bf 2.69} \\
{\bf ASR lattice m-vector} &  800 & 300 &{\bf 1.73} & (250,700)    & {\bf 1.93} & 300 / (300,300) & {\bf 1.62} \\ \hline
\end{tabular}
%\end{center}
\label{table:table3}
\end{table*}

\begin{figure}[!t]
%\centering
%\%begin{subfigure}[ht]{9.0cm}
%\centering
%\hspace*{-3.0cm}
\includegraphics[height=4.8cm,width=9.0cm]{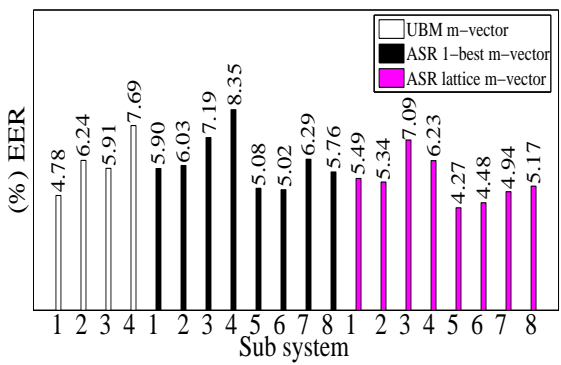}
%\vspace*{-0.5cm}
%\subcaption{\it With EFR}
%\label{fig:subsysEFR}
%\end{subfigure}
%\\[3.0ex]
%\begin{subfigure}[ht]{9.0cm}
%\centeri
%\hspace*{-3.0cm}
%\includegraphics[height=4.8cm,width=9.0cm]{figs/subSysPLDA.eps}
%\vspace*{-0.3cm}
%\subcaption{\it with PLDA}
%\label{fig:subsysPLDA}
%\end{subfigure}
\caption{\it EER value of each m-vector subsystem (cf. Table \ref{table:table3}) for task 7 of the NIST SRE 2008 core condition with EFR post-processing and scoring.}
\label{fig:subsys}
\end{figure}

\subsection{Comparison of the baseline and the m-vector systems}

Table~\ref{table:table3}(a)-(b) compares the performances of the optimal baseline systems with the proposed m-vector systems for the best parameter setup as found in earlier sections \ref{sec:opt_baseline_system} and \ref{sec:opt_m_vectorsize}, respectively,
for various post-processing and scoring techniques.
The proposed m-vector technique performs better than the baseline systems both with EFR and PLDA scoring, and the improvement is more important with the ASR-based approach (45-65\% rel. improvement) than with the UBM-based approach (10-20\% rel. improvement); in contrast, the performance with PLDA and EFR is comparable.
From Fig.~\ref{fig:subsys}, it can be observed that the EER of the individual m-vector sub-systems is much higher than that of the resulting fused system, showing that each sub-system extracts  relevant and complementary speaker information from the various parts of the MLLR super-vector.
%
%The proposed method is thus able to retrieve more speaker relevant information in the MLLR super-vector than the conventional way of directly using the full MLLR super-vector.
 
\subsection{Comparison of conventional and cascade post-processing}
As shown in Table \ref{table:table3}(c), the proposed cascade post-processing technique provides a lower EER than the conventional single post-processing (i.e., \emph{PLDA or LDA+EFR}) for MLLR-based systems in both the full as well as the m-vector cases, except in the UBM m-vector system configuration in which they are comparable.
This indicates that LDA as a first step followed by PLDA combines the benefits of both techniques. LDA projects the MLLR super-vectors or the m-vectors onto a low-dimensional discriminant vector space, and  PLDA then encounters less data sparsity in the parameter estimation in this reduced-dimensional space, yielding further gains and leading to 1.62\% EER for the ASR lattice m-vector configuration.

%\begin{table}[t]
%\begin{center}
%\caption{\it Performance of a standard i-vector system for the conventional single stage and proposed cascade post-processing method (shown in bold) on task 7 of NIST SRE 2008 core condition.}
%\begin{tabular}{l*{2}{l}r}\\ \hline
%  i-vector      & Post-processing        & \\
%   System       & (opt. parameters)      & \%EER    \\ \hline \hline
%      & LDA(300)+EFR          & 4.02  \\
%    & PLDA(350,400)          & 4.29  \\
%    & {\bf LDA(300)+PLDA(250,250)}   & {\bf 3.59}\\ \hline \hline
%\end{tabular}
%\label{table:table5}
%\end{center}
%\end{table}
%

In the presented cascade systems, PLDA was applied after selecting the optimal LDA projection dimension. Cascade post-processing was also performed with PCA or PPCA-NAP in the first stage and similarly showed  an improvement upon the single-step systems, but the cascade system with LDA pre-processing performed the best.
Further gains may be possible by optimizing the LDA, PCA or PPCA-NAP and PLDA parameters simultaneously.

\begin{table*}[t]
\begin{center}
\caption{\it Comparison of speaker verification performance of the proposed m-vector systems with the classical i-vector based system on task 7 of the NIST SRE 2008 core condition.}
\begin{tabular}{lr|rrr|rrr} \hline
 & & \multicolumn{3}{c|}{\bf (a) EFR post-processing} & \multicolumn{3}{c}{\bf (b) PLDA post-processing} \\
 System                                                            & i-/m-vector & Opt.  & & & Opt. dim  &  & \\ %\cline{2-3}
                                                                   &dim. & LDA proj. & \%EER      & MinDCF       & (spkr., chan.) & \%EER & MinDCF   \\ \hline \hline
{\bf (A)} i-vector                                                 &400  &300  & 4.02       & 0.0218       & (350,400)   & 4.29       & 0.0211       \\ 
{\bf (B)} UBM {\bf m-vector} (cf. Table \ref{table:table3})        &650  &400  & {\bf 3.81} & 0.0226       & (150,500)   & {\bf 3.50} & {\bf 0.0205} \\  
{\bf (C)} ASR 1-best {\bf m-vector} (cf. Table \ref{table:table3}) &750  &300  & {\bf 2.00} & {\bf 0.0130} & (500,500)   & {\bf 1.91} & {\bf 0.0131} \\
{\bf (D)} ASR lattice {\bf m-vector} (cf. Table \ref{table:table3})&800  &300  & {\bf 1.73} & {\bf 0.0134} & (250,700)   & {\bf 1.93} & {\bf 0.0134} \\&&&&&&&\\ %\hline
{\bf Late fusion}$^\ast$                                           &     &     &            &              &             &            &              \\ 
{\bf (A)+(B)}                                                      &-    &-    & {\bf 3.31} & {\bf 0.0199} & -           & {\bf 3.23} & {\bf 0.0166} \\
{\bf (A)+(C)}                                                      &-    &-    & 2.12       & {\bf 0.0122} & -           & 1.95       & {\bf 0.0128} \\ 
{\bf (A)+(D)}                                                      &-    &-    & {\bf 1.50} & {\bf 0.0122} & -           & {\bf 1.75} & {\bf 0.0121} \\&&&&&&&\\ %\hline 
{\bf Early fusion}                                                 &     &     &            &              &             &            &              \\
{\bf (A)+(B)}                                                      &1050 &500  & {\bf 2.97} & {\bf 0.0187} & (800,800)   & {\bf 2.78} & {\bf 0.0177} \\
{\bf (A)+(C)}                                                      &1150 &650  & 2.32       & 0.0146       &(750,1000)   & 2.97       & 0.0180       \\ 
{\bf (A)+(D)}                                                      &1200 &300  & 2.30       & 0.0145       & (1050,1000) & 2.83       & 0.0159       \\ \hline \hline
\end{tabular}\\
{$\ast$ Linear fusion}
\label{table:table6A}
\label{table:table6B}
\label{table:table6}
\end{center}
\end{table*}

\subsection{Combination of the i-vector and m-vector systems}

%The performance of a standard i-vector system with the conventional and with the proposed cascade post-processing techniques are presented in Table \ref{table:table5}.
%Only LDA followed by PLDA was considered, since i-vector dimension (i.e. $400$ in our configuration) is smaller compared to the full MLLR super-vectors.
%Speaker verification performance with PLDA and EFR is comparable, and the proposed cascade post-processing further reduces the EER, similar to the observation previously done for MLLR based systems.
  
Table \ref{table:table6} compares the i-vector-based speaker-verification system with the proposed m-vector systems
and presents the performance for their late fusion in the score domain and early fusion in the vector domain (i.e., concatenation of the i-vector to each m-vector).
The ASR-based m-vector performs better than the i-vector system with either EFR or PLDA, and the UBM-based system also shows promising performance.
The late fusion system further reduces EER and MinDCF for both EFR and PLDA in most of the cases. 
As expected, the lattice m-vector-based system shows a slightly lower error rate than the ASR $1$-best system.
In the case of the early fusion system involving ASR-based systems, EFR performs better than PLDA for post-processing and scoring.
This could be due to the fact that the early fusion systems result in larger dimensional vectors (i.e., m- plus i-vector size) and hence the PLDA system requires more training examples. 
%So, the proposed cascade system is able to further reduce either EER or MinDCF even compared to the best performance with EFR or PLDA especially for UBM and lattice based systems.
%Furthermore, for the other systems, the performance with cascade technique is comparable to their individual best system performance in early fusion domain. 
%Therefore, the proposed cascade post-processing is always beneficial to the early fusion system.
%This not only indicates the effectiveness of the proposed cascade post-processing technique but also that both i- and m-vector system contain complementary informations. 
%The performance of the i-vector system with either EFR or PLDA is comparable.

In the late fusion case, non-linear fusion of the scores of the various systems or sub-systems, may improve the speaker-verification performance compared to the results obtained with linear fusion. However, this requires additional data for training the parameters, and so it is kept as one of the perspectives of this work.

\subsection{Performance on NIST SRE 2010}
In this section, we compare the speaker-verification performance of the proposed m-vector systems with the i-vector on task 5 (telephone-telephone) of the NIST SRE 2010 core condition.
Only ASR based m-vectors are considered, because they showed significantly better performance than the full-MLLR and UBM-based m-vector systems on SRE 2008.
Table \ref{table:table7} compares the system performance for the various post-processing and scoring techniques proposed.
The m-vector size and the parameters used for LDA, PLDA and the cascade technique of the respective systems are the ones that were found to be optimal for the NIST SRE 2008, as presented in Tables~\ref{table:table3} and \ref{table:table6}.

\begin{table}[h]
\begin{center}
\caption{\it Comparison of the proposed m-vector system with a standard i-vector system on task 5 of the NIST SRE 2010 core condition for various post-processing techniques.}
\begin{subtable}{8.0cm}
\caption{{\bf EFR} and {\bf  PLDA } post-processing, scoring techniques}  
\hspace*{-0.2cm}
\begin{tabular}{lrr|rr} \hline
                                & \multicolumn{2}{c|}{\bf EFR}       & \multicolumn{2}{c}{\bf PLDA}  \\                                    
 System                       & \multicolumn{2}{c|}{\%EER MinDCF$^{\ast\ast}$} & \multicolumn{2}{c}{\%EER MinDCF$^{\ast\ast}$}\\
\hline \hline                                    
{\bf (A)} i-vector              & 3.69       & 0.7648       & 3.09       & 0.7406       \\
{\bf(C)} ASR 1-best m-vector    & {\bf 3.06} & {\bf 0.3399} & {\bf 2.83} & {\bf 0.2861} \\
{\bf (D)} ASR lattice m-vector  & {\bf 3.44} & {\bf 0.5042} & {\bf 2.92} & {\bf 0.3229} \\
                                &            &              &            &              \\
{\bf Late fusion$^\ast$}        &            &              &            &              \\
{\bf (A)+(C)}                   & {\bf 2.75} & {\bf 0.3172} & {\bf 2.29} & 0.3165       \\
{\bf (A)+(D)}                   & {\bf 2.35} & {\bf 0.2832} & {\bf 2.72} & {\bf 0.3144} \\
                                &            &              &            &              \\
{\bf Early fusion}              &            &              &            &              \\
{\bf (A)+(C)}                   & {\bf 2.59} & 0.5941       & {\bf 2.61} & 0.4589       \\
{\bf (A)+(D)}                   & {\bf 2.78} & 0.4517       & {\bf 2.84} & 0.4411       \\ 
\hline \hline
\end{tabular}
%\label{table:table7} \\
{\scriptsize $^\ast$  Linear fusion; $^{\ast\ast}$ MinDCF as  per NIST SRE 2010}
%  (see ref. Tables \ref{table:table6A} \& \ref{table:table6B} for opt. parameters).
\label{table:table7A}
\end{subtable}
\\[2.8ex]
\begin{subtable}{7.0cm}
\hspace*{-2.0cm}
\caption{{\bf Cascade} (LDA+PLDA) post-processing technique} % \\[1.0ex]
\hspace*{-0.2cm}
\begin{tabular}{l*{4}{c}r}\\ \hline
 System                       & \%EER & MinDCF \\ \hline  \hline
i-vector                      & {\bf 3.01} & {\bf 0.5977}    \\
ASR 1-best m-vector           & {\bf 2.46} & {\bf 0.2776}    \\
ASR lattice m-vector          & {\bf 2.08} & {\bf 0.2436}   \\ \hline \hline
\end{tabular}
\label{table:table7B}
%{\scriptsize (see ref. Tables \ref{table:table4} \& \ref{table:table5} for opt. parameters).}

\end{subtable}
\label{table:table7}
\end{center}
\end{table}

The proposed m-vector system yields better performance than the i-vector in terms of both EER and MinDCF for the EFR and PLDA post-processing and scoring techniques, similar to the result on the NIST SRE 2008 in Table \ref{table:table3}. 
Further, early fusion in the feature domain as well as late fusion in the score domain of both the m- and i-vectors reduces the EER in all configurations, and also reduces the MinDCF for late fusion in the EFR configuration. This confirms the complementarity between the m- \& i-vectors.
Additionally, the proposed cascade post-processing technique shows significantly lower EER and MinDCF values compared to the conventional single post-processing (i.e., PLDA or LDA+EFR) of the respective systems.
Combining cascade post-processing and fusion between the i- and m-vector systems would be a natural extension of this work.

It is to be noted that on the same dataset (i.e., male speakers trials of the NIST SRE 2010 condition 5), Scheffer et al. reported 0.432 MinDCF and 3.29\% EER for their best MLLR configuration~\cite{Nicolas_Scheffer2011}.
For the male trials of the NIST SRE 2010 extended task with significantly more trials and training data, Cumani et al. reported 0.470 MinDCF and 4.15\% EER for a MLLR PLDA system~\cite{Cumani2012}.
Our MLLR m-vector systems compare favorably to these results.
Both studies~\cite{Nicolas_Scheffer2011,Cumani2012} also presented a baseline system with NAP-compensated MLLR vectors and an SVM back-end, which stood significantly behind the other systems; and we therefore did not consider an SVM-based system in our work.

\section{Conclusions}
\label{sec:conclusion}
In this paper, we addressed speaker verification (SV) using m-vectors for speaker representation, in which m-vectors are extracted by a uniform segmentation of the speaker MLLR super-vector using an overlapping window.
We compared two main techniques for session-variability compensation and scoring, namely eigen factor radial (EFR) and probabilistic linear discriminant analysis (PLDA), 
and we proposed  a cascade post-processing technique for speaker verification using the m-vector or MLLR super-vector.
The system performances were demonstrated for male speakers for a standard task of the SRE 2008 and SRE 2010 core condition.

Our experiments show that the proposed m-vector approach performs significantly better than the baseline approach of directly processing the full MLLR super-vector in all configurations.
This indicates that it is able to retrieve more speaker-relevant information from the speaker MLLR super-vector than the conventional approach.
The most obvious explanation is related to the data sparsity issue that arises when estimating the LDA or PLDA projection matrices with the full MLLR super-vectors. The m-vector approach may be sub-optimal in the sense that each m-vector only represents a subset of the initial MLLR super-vector. A single m-vector sub-system is indeed less efficient than the baseline system. However, the fusion of the different m-vector sub-systems enables a more precise estimation of the projection space and, finally, a more robust and better-performing system. 

The same explanation can be proposed with the cascade post-processing technique, which yields reductions in the SV error rates compared to the standalone PLDA or LDA+EFR systems.
In this technique, the m-vector or MLLR super-vector is first projected onto a low-dimensional discriminant vector space generated by LDA. Then, PLDA is applied on the reduced-dimensional vector for session-variability compensation and scoring.
This two-stages process combines the advantages of both techniques and reduces the possible data-sparsity problem at the  PLDA stage compared to the direct use of the full dimension vector.

As expected, the improvement in the SV performance with ASR-based systems over that of the UBM-based system shows the effectiveness of integrating phonetic knowledge into the MLLR transformation. The ASR lattice-based method performed slightly better than the $1$-best transcription, because it was more robust to transcription errors.

Finally, we compared the performance of the proposed m-vector with a standard i-vector system associated with EFR and PLDA post-processing.
The proposed ASR-based m-vector system showed consistently better performance than the i-vector system.
We also considered the late fusion in the score domain and the early fusion in the vector domain for m- and i-vector systems.
Both fusion cases provided further improvement of the SV performance compared to the individual systems, showing that the m- and i-vectors contain complementary speaker-relevant informations.

\section*{Acknowledgments}
This work was realized through the QUAERO Program and the QCOMPERE project, which were funded by OSEO (French State agency for innovation, now Bpifrance) and ANR (French national research agency), respectively.

% -------------------------------------------------------------------------
\bibliographystyle{IEEEbib}
\bibliography{References}

@Article{Mahalanobis36,
	author={P. C. Mahalanobis},
	title={{O}n the {G}eneralised {D}istance in {S}tatistics},
	journal={Proc. of the National Institute of Sciences of India},
        volume={2},
	pages={49-55},
	year={1936}
}


@Article{MacQueen67,
	author={J. MacQueen},
	title={{S}ome {M}ethods for {C}lassification and {A}nalysis of {M}ultivariate {O}bservations},
	journal={Proc. Fifth Berkeley Symp. on Math. Statist. and Prob. (Univ. of Calif. Press)},
	volume={1},
	pages={281-297},
	year={1967}
}

@Article{ross74,
	author={M. J. Ross and  H. L. Shaffer and A. Cohen and R. Freudberg and H. J.Manley},
	title={{A}verage {M}agnitude {D}ifference {F}unction {P}itch {E}xtractor},
	journal={IEEE Trans. Acoust. Speech Signal Processing},
	volume={22},
	pages={353-362},
	year={1974}
}

@INPROCEEDINGS{Makhoul75,
	author = {J. Makhoul},
	title = {{L}inear {P}rediction: {A} {T}utorial {R}eview},
	booktitle = {Proceedings of the IEEE},
	volume={63},
	pages={561-580},
	year={1975}
}

@Article{Sambur76,
	author = {M. R. Sambur},
	title = {{S}peaker {R}ecognition using {O}rthogonal {L}inear {P}rediction},
	journal={IEEE Trans. Acoust. Speech Signal Processing},
	volume={24},
	pages={283-289},
	year={1976}
}
	
@Article{Markel77,
	author = {J. D. Markel and B. T. Oshika and Jr. A. M. Gray},
	title = {{L}ong-term {F}eature {A}veraging for {S}peaker {R}ecognition},
        journal={IEEE Trans. Acoust. Speech Signal Processing},
	volume={25},
	pages={330-337},
	year={1977}
}

@Article{dempster77,
  author={A. Dempster and N. Laird and D. Rubin},
  title={{M}aximum {L}ikelihood from {I}ncomplete {D}ata via {EM} {A}lgorithm},
  journal={J. Roy. statist. Soc.},
  volume={39},
  pages={1-38},
  year={1977}
   }

@Article{Davis80,
	author={S. B. Davis and P. Mermelstein},
	title={{C}omparison of {P}arametric {R}epresentations for {M}onosyllabic {W}ord {R}ecognition in {C}ontinuously {S}poken {S}entences},
	journal={IEEE Trans. Acoust. Speech Signal Processing},
	volume={28},
	pages={357-366},
	year={1980}
}

@Article{LINDE80,
	author={Y. Linde and A. Buzo and  R. M. GRAY},
        title={{A}n {A}lgorithm for {V}ector {Q}uantizer {D}esign},
	journal={IEEE Trans. Communications},
	volume={28},
	pages={84-94},
	year={1980}
}

@Article{Furui81,
	author={S. Furui},
	title={{C}epstral {A}nalysis {T}echnique for {A}utomatic {S}peaker {V}erification},
	journal={IEEE Trans. on  Acoust. Speech Signal Processing},
	volume={29},
	pages={254-272},
	year={1981}
}

@Article{Rabiner89,
	author={L. R. Rabiner},
	title={{A} {T}utorial on {H}idden {M}arkov {M}odels and {S}elected {A}pplications in {S}peech {R}ecognition},
	journal={Proc. of the IEEE},
	volume={77},
	pages={257-285},
	year={1989}
}



@Article{Hermansky90,
	author={H. Hermansky},
	 title={{P}erceptual {L}inear {P}redictive ({PLP}) {A}nalysis of {S}peech},
	 journal={J. Acoust. Soc. Am.},
	 volume={87},
	 pages={1738-1752},
	 year={1990}
}

@Article{Higgin91,
author="A. Higgins and L. Bahler and J. Porter",
title="{S}peaker {V}erification using {R}andomized {P}hrase {P}rompting",
journal="Digital Signal Processing",
volume="1",
 pages="89-106",
year="1991"	 
}

@INPROCEEDINGS{reynolds91,
	author = {D. A. Reynolds and L. P. Heck},
	title = {{I}ntegration of {S}peaker and {S}peech {R}ecognition {S}ystems},
         booktitle = {Proc. of IEEE Int. Conf. Acoust. Speech Signal Processing (ICASSP)},
	 pages = {869-872},
	year = {1991}
}


@INPROCEEDINGS{Rosenberg92,
	author = {A. E. Rosenberg and J. DeLong and C\-H Lee and B\-H Jaung and F. K. Soong},
	title = {{T}he {U}se of {C}ohort {N}ormalized {S}cores for {S}peaker {V}erification},
	booktitle = { Proc. of Int. Conf. Spoken Language Processing (ICSLP)},
	pages = {599-602},
	year = {1992}
}

@Article{Hermanksy94,
       author = {H. Hermanksy and N. Morgan},
       title = {{R}ASTA {P}rocessing of {S}peech},
       journal={IEEE Trans. on Speech and Audio Processing},
       volume={2},
       pages={578-589},
       year={1994}
}

@Article{Gauvain94,
	author = {J.-L. Gauvain and C.-H. Lee},
	title = {{M}aximum a {P}osteriori {E}stimation for {M}ultivariate {G}aussian {M}ixture {O}bservations of {M}arkov {C}hains},
	journal={IEEE Trans. on Speech and Audio Processing},
	volume={2},
	pages={291-298},
	 year={1994}
}


@Article{reynold95,
  author="D. A. Reynolds",
  title="{S}peaker {I}dentification and {V}erification using {G}aussian {M}ixture {S}peaker {M}odels",
  journal="Speech Communication",
  volume="17",
  pages="91-108",
  year="1995"
}

@Article{Leggeter95,
author="C. Leggetter and P. Woodland",
title="{M}aximum {L}ikelihood {L}inear {R}egression for {S}peaker {A}daptation of {HMM}s",
journal="Computer Speech and Language",
volume="9",
pages="171-186",
year="1995"	
}
@Article{Vapnik95,
	author={C. Cortes and V. Vapnik},
	title={{S}upport {V}ector {N}etworks},
	journal={Machine Learning},
	volume={20},
	pages={273-297},
	year={1995}
}

@INPROCEEDINGS{Talkin95,
	author = {D. Talkin},
        title = {{A} {R}obust {A}lgorithm for {P}itch {T}racking ({RAPT})},
	booktitle = {in Speech coding and synthesis (Elsevier, ed.)},
	pages = {495-518},
        year = {1995}
}

@INPROCEEDINGS{leerose96,
 author = {Li Lee and Richard C. Rose},
 title = {{S}peaker {N}ormalization using {E}fficient {E}requency {W}arping {P}rocedures},
 booktitle = {Proc. of IEEE Int. Conf. Acoust. Speech Signal Processing (ICASSP)},
 pages = {353-356},
 year = {1996}
}



@INPROCEEDINGS{Rosenberg96,
	 author = {A. E. Rosenberg and S. Parthasarathy},
	 title = {{S}peaker {b}ackground {m}odels for {c}onnected {d}igit {p}assword {s}peaker {V}erification},
	 booktitle = {Proc. of IEEE Int. Conf. Acoust. Speech Signal Processing (ICASSP)},
	 pages = {81-84},
	 year = {1996}
}




@INPROCEEDINGS{LHeck97,
	  author = {L. Heck and M. Weintraub},
	  title = {{H}andset-dependent {B}ackground {M}odels for {R}obust {T}ext-{I}ndependent {S}peaker {R}ecognition},
          booktitle = {Proc. of IEEE Int. Conf. Acoust. Speech Signal Processing (ICASSP)},
	  pages = {1071-1074},
	  year = {1997}
}


@INPROCEEDINGS{DET97,
	   author = {A. Martin and G. Doddington and T. Kamm and M. Ordowskiand and M. Przybocki},
	   title = {{T}he {D}ET {C}urve in {A}ssessment of {D}etection {T}ask {P}erformance},
	   booktitle = {Proc. of Eur. Conf. Speech Commun. and Tech. (Eurospeech)},
	   pages = {1895-1898},
	   year = {1997}
}  


@INPROCEEDINGS{Ariyaeeinia97,
	author= {A. M. Ariyaeeinia and P. Sivakumaran},
	title = {{A}nalysis and {C}omparison of {S}core {N}ormalization {M}ethods for {T}ext {D}ependent {S}peaker {V}erification},
	booktitle = {Proc. of Eur. Conf. Speech Commun. and Tech. (Eurospeech)},
        pages = {1379-1382},
	year = {1997}
}


@article{Lee-Rose98,
      author="L. Lee and R. Rose",
      title="{F}requency {W}arping {A}pproach  to {S}peaker {N}ormalization",
      journal="IEEE Trans. on Speech and Audio Processing",
      volume={6},
      pages={49-59},
      year={1998}
}



@INPROCEEDINGS{reycls98,
	author= {D. A. Reynolds and et. al},
	title = {{B}lind {C}lustering of {S}peech {U}tterances {B}ased on {S}peaker and {L}anguage {C}haracteristics},
	booktitle = {Proc. of Int. Conf. Spoken Language Processing (ICSLP)},
	pages = {3193-3196},
	year={1998}
}

@INPROCEEDINGS{Hermansky98,
	author={S. Van Vuuren and H. Hermansky},
	title={{O}n the {I}mportance of {C}omponents of the {M}odulation {S}pectrum for {S}peaker {V}erification},
        booktitle = {Proc. of Int. Conf. Spoken Language Processing (ICSLP)},
	pages = {3205-3208},
	year = {1998}
}


@Article{Pellom98,
    author={B. L. Pellom and J. H. L. Hansen},
    title={{A}n {E}fficient {S}coring {A}lgorithm for {G}aussian {M}ixture {M}odel {B}ased {S}peaker {I}dentification},
    journal={IEEE Signal Proc. Lett.},
    volume={5},
    pages={281-284},
    year={1998}
}

@Article{HLDA98,
	author= {N. Kumar and A.G. Andreou},
	title={{H}eteroscedastic {D}iscriminant {A}nalysis and {R}educed {R}ank {HMMs} for {I}mproved {S}peech {R}ecognition},
	journal={Speech Communication},
	volume={26},
	pages={283-297},
	year={1998}
}

@INPROCEEDINGS{Isobe99,
	author= {T. Isobe and J. Takahashi},
	title = {{A} {N}ew {C}ohort {N}ormalization using {L}ocal {A}coustic {I}nformation for {S}peaker {V}erification},
	booktitle = {Proc. of IEEE Int. Conf. Acoust. Speech Signal Processing (ICASSP)},
        pages = {841-844},
	year={1999}
}

@INPROCEEDINGS{McLaugn99,
     author={J. McLaughlin and D. A. Reynolds and T. Gleason},
     title={{A} {S}tudy of {C}omputation {S}peed-ups of the {GMM-UBM} {S}peaker {R}ecognition {S}ystem},
     booktitle = {Proc. of Eur. Conf. Speech Commun. and Tech. (Eurospeech)},
     pages={1215-1218},
     year={1999}
}

@INPROCEEDINGS{Markowitz99,
	author={J. Markowitz},
       title={{I}eri and {O}ggi {D}omani: {S}peaker {R}ecognition {Y}esterday, {T}oday and tomorrow},
       booktitle = {Proc. of COST250 Workshop on Speacker Recognition in Telephony},
       year={1999}
}



@Article{kuhn2000,
author="R. Kuhn and J.-C. Junqua and P. Nguyen and N. Niedzielski",
title="{R}apid {S}peaker {A}daptation in {E}igenvoice {S}pace",
journal="IEEE Trans. on Speech and Audio Processing",
volume="8 (6)",
pages="695-707",
year="2000"	
}	  

@Article{reynold00,
  author="D. A. Reynolds and T. F. Quatieri and R. B. Dunn",
  title="{S}peaker {V}erification using {A}dapted {G}aussian {M}ixture {M}odels",
  journal="Digital Signal Processing",
  volume="10",
  pages="19-41",
 year="2000"	  
}

@Article{Tnorm00,
 author="R. Auckenthaler and M. Carey and H. Lloyd-Thomas",
 title="{S}core {N}ormalization for {T}ext-{I}ndependent {S}peaker {V}erification {S}ystem",
 journal="Digital Signal Processing",
 Volume="10",
 pages="42-54",
 year="2000"
 }


@INPROCEEDINGS{Wagner00,
     author= {D. Tran and M. Wagner},
     title = {{A} {P}roposed {L}ikelihood {T}ransformation for {S}peaker {V}erification},
     booktitle = {Proc. of IEEE Int. Conf. Acoust. Speech Signal Processing (ICASSP)},
     pages = {1069-1072},
     year={2000}
}


@Article{kishore00,
  author="S. P. Kishore and B. Yegnanarayana",
  title="{S}peaker {V}erification: {M}inimizing the {C}hannel {E}ffects using {A}utoassociative {N}eural {N}etwork {M}odels",
  journal="ICASSP",
  year="2000",
  pages="1101-1104"
}



@INPROCEEDINGS{Heck00,
	author= {R. Teunen and B. Shahshahani and L. Heck},
	title = {{A} {M}odel-based {T}ransformational {A}pproach to {R}obust {S}peaker {R}ecognition},
	booktitle = {Proc. of Int. Conf. Spoken Language Processing (ICSLP)},
        pages = {495-498},
        year={2000}
}

@INPROCEEDINGS{paliwal00,
	author= {B. Wildermoth and K. K. Paliwal},
	title = {{U}se of {V}oicing and {P}itch {I}nformation for {S}peaker {R}ecognition},
	booktitle = {Proc. 8th Australian Intern. Conf. Speech Science and Technology},
	pages = {324-328},
       year={2000}

}


@INPROCEEDINGS{Gadde2000,
	author= {V. R. R. Gadde},
	title = {{M}odeling {W}ord {D}uration},
        booktitle = {Proc. of Int. Conf. Spoken Language Processing (ICSLP)},
	pages = {601-604},
        year={2000}
}

@INPROCEEDINGS{Wan2000,
	author= {V. Wan and W. M. Campbell},
	title = {{S}upport {V}ector {M}achines for {S}peaker {V}erification and {I}dentification},
	booktitle = {Proc. IEEE Signal Processing Society Workshop Neural Networks},
	pages = {775-784},
	year={2000}
}

@INPROCEEDINGS{padmanabhan_00,
author = {M. Padmanabhan and G. Saon and G. Zweig},
title = {{L}attice-{B}ased {U}nsupervised {MLLR} for {S}peaker {A}daptation},
booktitle = {Proc. of the ISCA ITRW ASR2000},
pages = "128-131",
year = {2000},
}

@INPROCEEDINGS{Doddington2001,
        author= {G. Doddington},
        title = {{S}peaker {R}ecognition {B}ased on {I}diolectal {D}ifferences {B}etween {S}peakers},
        booktitle = {Proc. of Eur. Conf. Speech Commun. and Tech. (Eurospeech)},
        pages = {2521-2524},
        year={2001}
}  

@INPROCEEDINGS{uebel_01,
author = {L.F. Uebel and P.C. Woodland},
title = {{I}mprovements in {L}inear {T}ranformation based {S}peaker {A}daptation},
booktitle = {Proc. of IEEE Int. Conf. Acoust. Speech Signal Processing (ICASSP)},
pages={49-52},
year = {2001}
}

@INPROCEEDINGS{Sturim01,
        author= {D. Sturim and others},
        title = {{S}peaker {I}ndexing in {L}arge {A}udio {D}atabases using {A}nchor {M}odels},
	booktitle = {Proc. of ICASSP},
	 pages = {429-432},
        year={2001}
}

@INPROCEEDINGS{Auck01,
       author={R. Auckenthaler and J. S. Masion},
       title={{G}aussian {S}election {a}pplied to {T}ext-{I}ndependent {S}peaker {V}erification},
       booktitle = {Proc. of Odyssey Speaker and Language Recognition Workshop},
       pages={83-88},
       year={2001}
}

@INPROCEEDINGS{Pelecanos01,
       author={J. Pelecanos and S. Sridharan},
       title={{F}eature {W}arping for {R}obust {S}peaker {V}erification},
       booktitle = {Proc. of Odyssey Speaker and Language Recognition Workshop},
       pages={213-218},
       year= {2001}
}


@INPROCEEDINGS{Wang2001,
        author= {J. C. N. Wang and W. H. Tsai and L. S. Lee},
        title = {{E}igen-{MLLR} {C}oefficients as {N}ew {F}eature {P}arameters for {S}peaker {I}dentification},
        booktitle = {Proc. of Eur. Conf. Speech Commun. and Tech. (Eurospeech)},
	pages = {1385-1388},
        year={2001}
}



@INPROCEEDINGS{Glds_ker2002,
	author={W. M. Campbell},
	title={{G}eneralized {L}inear {D}iscriminant {S}equence {K}ernels for {S}peaker {R}ecognition},
	booktitle = {Proc. of IEEE Int. Conf. Acoust. Speech Signal Processing (ICASSP)},
	pages = {161-164},
	year={2002}
}


@INPROCEEDINGS{ReynoldsFmap,
	author= {D.A. Reynolds},
	title = {{C}hannel {R}obust {S}peaker {V}erification via {F}eature {M}apping},
	booktitle = {Proc. of IEEE Int. Conf. Acoust. Speech Signal Processing (ICASSP)},
	pages = {6-10},
        year={2003}
}


@INPROCEEDINGS{Tkinnun03,
    author={T. Kinnunen and E. Karpov and P. Franti},
    title={{A} {S}peaker {P}runing {A}lgorithm for {R}eal-{T}ime {S}peaker {I}dentification},
    booktitle = {Proc. Audio- and Video-Based Biometric Authentication},
    pages={639-646},
        year={2003}
}




@Article{nisteval04,
	author="{T}he {E}valuation {P}lan of NIST 2004 {S}peaker {R}ecognition {C}ampaign.
	http://www.itl.nist.gov/iad/mig//tests/sre/2004/SRE-04\_evalplan{-}v1a.pdf"
}

@INPROCEEDINGS{bonastre04,
        author = {J. F. Bonastre and N. Scheffer and C. Fredouille and D. Matrouf},
        title = {{N}IST'04 {S}peaker {R}ecognition {E}valuation {C}ampaign: {N}ew {LIA} {S}peaker {D}etection {P}lateform based on {ALIZE} {T}oolkit},
        booktitle = {Proc. of NIST 2004 Speaker Recognition Workshop},
        year = {2004},
}

@INPROCEEDINGS{Binma04,
	author = {B. Ma and H. Meng},
	title = {{E}nglish-{C}hinese {b}ilingual {t}ext-independent {s}peaker {v}erification},
	booktitle = {Proc. of IEEE Int. Conf. Acoust. Speech Signal Processing (ICASSP)},
	year={2004}
}

@INPROCEEDINGS{Leung04,
	author = {K. Y. Leung and M. W. Mak and S. Y. Kung},
	title = {{A}pplying {a}rticulatory {f}eatures to {t}elephone-based {s}peaker {v}erification},
        booktitle = {Proc. of IEEE Int. Conf. Acoust. Speech Signal Processing (ICASSP)},
	year={2004}
}

@INPROCEEDINGS{Kwong04,
	author = {Q.Y. Hong and S. Kwong},
	title = {{D}iscriminative {t}raining for {s}peaker {i}dentification based on {m}aximum {m}odel {d}istance {a}lgorithm},
        booktitle = {Proc. of IEEE Int. Conf. Acoust. Speech Signal Processing (ICASSP)},
	year={2004}
}
@INPROCEEDINGS{Prasad05,
       author = {R. Prasad and others},
       title = {{T}he 2004 {BBN}/{LIMSI} {20xRT} {E}nglish {C}onversational {T}elephone {S}peech {R}ecognition {S}ystem}, 
       booktitle = {Proc. of INTERSPEECH},
       pages={1645-1648},
       year={2005}
}



@INPROCEEDINGS{Louradour05,
	author = {J. Louradour and K. Daoudi and R. Andre-Obrecht},
	title = {{D}iscriminative {P}ower of {T}ransient {F}rames in {S}peaker {R}ecognition},
	booktitle = {Proc. of IEEE Int. Conf. Acoust. Speech Signal Processing (ICASSP)},
	pages={613-616},
	year={2005}
}



@inproceedings{Stolcke05,
	Author = {A. Stolcke and L. Ferrer and S. Kajarekar and E. Shriberg and A. Venkataraman},
	Booktitle = {Proc. of EUROSPEECH},
	Date-Modified = {2014-02-03 08:41:06 +0000},
	Month = {September},
	Pages = {2425-2428},
	Title = {{MLLR} {T}ransforms as {F}eatures in {S}peaker {R}ecognition},
	Year = {2005}}


@Article{kenny05,
author="P. Kenny",
title="{J}oint {F}actor {A}nalysis of {S}peaker and {S}ession {V}ariability : {T}heory and {A}lgorithms",
journal="Technical report CRIM-06/08-13  Montreal, CRIM",
year="2005"
}



@INPROCEEDINGS{AT_norm,
	author= {D. E. Sturim and D.A. Reynolds},
	title = {{S}peaker {A}daptive {C}ohort {S}election for {T}norm in {T}ext-independent {S}peaker {V}erification},
        booktitle = {Proc. of IEEE Int. Conf. Acoust. Speech Signal Processing (ICASSP)},
	pages = {741-744},
        year={2005}
}	


@INPROCEEDINGS{datadriven05,
	author= {M. Mason and R. Vogt and B. Baker and S. Sridharan},
	title = {{D}ata-{D}riven {C}lustering for {B}lind {F}eature {M}apping in {S}peaker {V}erification},
	booktitle = {Proc. of Interspeech},
         pages = {3109-3112},
        year={2005}
}


@INPROCEEDINGS{ZTnorm05,
	author= {R. Vogt and B. Baker and S. Sridharan},
	title = {{M}odeling {S}ession {V}ariability in {T}ext-{I}ndependent {S}peaker {V}erification},
	booktitle = {Proc. of Eur. Conf. Speech Commun. and Tech. (Eurospeech)},
	pages = {3117-3120},
	year={2005}
}

@INPROCEEDINGS{Adami05,
	author= {André G. Adami},
        title = {{P}rosodic {M}odeling for {S}peaker {R}ecognition based on {S}ub-band {E}nergy {T}emporal {T}rajectories},
	booktitle = {Proc. of IEEE Int. Conf. Acoust. Speech Signal Processing (ICASSP)},
	 pages = {189 - 192},
        year={2005}
}

@INPROCEEDINGS{NAP2005,
	author= {A. Solomonoff and W. M. Campbell and I. Boardman},
	title = {{A}dvances in {C}hannel {C}ompensation for {SVM} {S}peaker {R}ecognition},
	booktitle = {Proc. of IEEE Int. Conf. Acoust. Speech Signal Processing (ICASSP)},
	pages = {629 - 632},
	year={2005}
}

@INPROCEEDINGS{Goldberger05adistance,
author = {J. Goldberger and H. Aronowitz},
title = {{A} {D}istance {M}easure {B}etween {GMM}s {B}ased on the {U}nscented {T}ransform and its {A}pplication to {S}peaker {R}ecognition},
booktitle = {in Proc. of Interspeech},
pages = {1985-1989},
year = {2005}
}



@INPROCEEDINGS{krause06,
	author = {N. Krause and R. Gazit},
	title ={{SVM}-based {S}peaker {C}lassification in the {GMM} {M}odels {S}pace},
	booktitle = {Proc. of Odyssey Speaker and Language Recognition Workshop},
	year ={2006}
}


@INPROCEEDINGS{Noor06,
	  author = { E. Noor and H. Aronowitz},
	  title={{E}fficient {L}anguage {I}dentification using {A}nchor {M}odels and {S}upport {V}ector {M}achines},    
	  booktitle = {Proc. of Odyssey Speaker and Language Recognition Workshop},
	  year ={2006}
}



@INPROCEEDINGS{Binma06,
	author = {B. Ma and D. Zhu and R. Tong and H. Li},
	title={{S}peaker {C}luster {B}ased {GMM} {T}okenization for {S}peaker {R}ecognition},
	booktitle = {Proc. of Interspeech},
        year ={2006},
}


@INPROCEEDINGS{Hazen06,
  author="R. Woo and A. Park and T. J. Hazen",
  title="{T}he {MIT} {M}obile {D}evice {S}peaker {V}erification {C}orpus: {D}ata {C}ollection and {P}reliminary {E}xperiments",
  booktitle="Proc. of Odyssey Speaker and Language Recognition Workshop",
  pages="1-6",
  year="2006"
}


@Article{Campbell2006,
	author={W. M. Campbell and D. E. Sturim and D. A. Reynolds},
	title={{S}upport {V}ector {M}achines using {GMM} {S}upervectors for {S}peaker {V}erification},
	 journal={IEEE Signal Process. Lett.},
	 volume={13},
	 pages={308-311},
	 year={2006}
}


@Article{Mami06,
	author="Y. Mami and D. Charlet",
	title="{S}peaker {R}ecognition by {L}ocation in the {S}pace of {R}eference {S}peakers",
	journal="Speech Communication",
	volume="48",
	pages="127-141",
	year="2006"
}

@Article{Skosan06,
	author={M. Skosan and D. Mashao},
	title={{M}odified {S}egmental {H}istogram {E}qualization for {R}obust {S}peaker {V}erification},
	journal={Pattern Recogn. Lett.},
        volume={27},
        pages={479–486},
        year={2006}
}	


@INPROCEEDINGS{hatch06,
   author = {A. Hatch and S. Kajarekar and A. Stolcke},
   title={{W}ithin-{C}lass {C}ovariance {N}ormalization for {SVM}-{B}ased {S}peaker {R}ecognition},
   booktitle = {Proc. of ICSLP},
   pages = {1471-1474},
   year ={2006}
}


@Article{htkbook,
  author="S. Young and D. Kershaw and J. Odell and V. Valtchev and P. Woodland and et al.",
  title="{HTK} {B}ook",
  journal={Copyright 2001-2006 Cambridge University Engineering Department}
   }


@Article{Castro2007,
      author="D. R. Castro and et. al",
      title="{S}peaker {V}erification using {S}peaker- and {T}est-dependent {F}ast {S}core {N}ormalization",
      journal="Pattern Recogn. Lett.",
      volume="28",
      pages="90-98",
      year="2007"
}

@INPROCEEDINGS{Charbuillet07,
	author={C. Charbuillet and B. Gas and M. Chetouani and J. L. Zarader},
	title={{C}omplementary {F}eatures for {S}peaker {V}erification based on {G}enetic {A}lgorithms},
        booktitle = {Proc. of IEEE Int. Conf. Acoust. Speech Signal Processing (ICASSP)},
	pages={285-288},
	year = {2007}
}

@INPROCEEDINGS{Xugang07,
	author={X. Lu and J. Dang},
	 title={{P}hysiological {F}eature {E}xtraction for {T}ext {I}ndependent {S}peaker {I}dentification using {N}on-{U}niform {S}ub-band {P}rocessing},
        booktitle = {Proc. of IEEE Int. Conf. Acoust. Speech Signal Processing (ICASSP)},
	pages={461-464},
        year = {2007}
}

@INPROCEEDINGS{Saeidi07,
	author={R. Saeidi and H. R. Sadegh Mohammadi and R. D. Rodman and T Kinnunen},
	title={{A} {N}ew {S}egmentation {A}lgorithm {C}ombined with {T}ransient {F}rames {P}ower for {T}ext {I}ndependent {S}peaker {V}erification},
	booktitle = {Proc. of IEEE Int. Conf. Acoust. Speech Signal Processing (ICASSP)},
	pages={305-308},
	year = {2007}
}
		
@INPROCEEDINGS{Tur2007,
	author={G. Tur and E. Shriberg and A. Stolcke and S. Kajarekar},
	title={{D}uration and {P}ronunciation {C}onditioned {L}exical {M}odeling for {S}peaker {S}erification},
        booktitle = {Proc. of Interspeech},
	pages={2049-2052},
	year = {2007}
}

@Article{HLDA_feat07,
	author={L. Burget and  P. Matejka and  O. Glembek and P. Schwarz and J. Cernocky},
        title = {{A}nalysis of {F}eature {E}xtraction and {C}hannel {C}ompensation in {GMM} {S}peaker {R}ecognition {S}ystem},
	journal={IEEE Trans. on Audio, Speech, Lang. Process.},
	volume={15},
	pages={1979-1986},
	year={2007}
}

@Article{fusion_2007,
	 author={N. Brummer and L. Burget and J. Cernocky and others},
	 title = {{F}usion of {H}eterogeneou {S}peaker {R}ecognition {S}ystems in the {STBU} {S}ubmission for the {NIST} {S}peaker {R}ecognition {E}valuation 2006},
	 journal={IEEE Trans. on Audio, Speech, and Lang. Process.},
	 volume={15},
	 pages={2072-2084},
	 year={2007}
}

@INPROCEEDINGS{Hazen97,
       author= {T. J. Hazen and J. R. Glass},
       title = {{A} {C}omparison of {N}ovel {T}echniques for {I}nstantaneous {S}peaker {A}daptation},
       booktitle = {Proc. of Eur. Conf. Speech Commun. and Tech. (Eurospeech)},
       pages = {2047-2050},
       year = {1997}
}


@INPROCEEDINGS{ferras-icassp08,
	author = {M. Ferras and others},
        title = {{C}onstrained {MLLR} for {S}peaker {R}ecognition},
	booktitle = {Proc. of IEEE Int. Conf. Acoust. Speech Signal Processing (ICASSP)},
	pages = {53-56},
	year = {2007}
}


	       


@inproceedings{akhil-interspeech2008,
      author = {P. T. Akhil and S. P. Rath and S. Umesh and D. R. Sanand},
      title = {{A} {C}omputationally {E}fficient {A}pproach to {W}arp {F}actor {E}stimation in {VTLN} Using {EM} {A}lgorithm and {S}ufficient {S}tatistics},
      booktitle = {Proc. of Interspeech},
      pages = {1713-1716},
      year = {2008}
}


@inproceedings{sanand-interspeech2008,
      author={D. R. Sanand and S. Umesh},
      title={{S}tudy of {J}acobian {C}ompensation Using {L}inear {T}ransformation of {C}onventional {MFCC} for {VTLN}},
      booktitle={Proc. of Interspeech},
      pages={1233-1236},
      year={2008}
}
@inproceedings{KaramIcassp2008,
	author={Z. N. Karam and W. M. Campbell},
	title={{A} {M}ulti-class {MLLR} {K}ernel for {SVM} {S}peaker {R}ecognition},
	 booktitle={Proc. of ICASSP},
         pages={4117-4120},
	 year = {2008}
      
}

@inproceedings{Fousek2008,
        author={P. Fousek and L. Lamel and J. L. Gauvain},
        title={{T}ranscribing {B}roadcast {D}ata using {MLP} {F}eatures},
        booktitle = {Proc. of INTERSPEECH},
        pages={1433-1436},
        year={2008}
}


@inproceedings{Yossi09,
	author = {Y. Bar-Yosef and Y. Bistritz},
        title = {{A}daptive {I}ndividual {B}ackground {M}odel for {S}peaker {V}erification},
        booktitle = {Proc. of Interspeech},
	pages = {1271-1274},
	year = {2009}
}	

@Article{Vraj09,
	author="V. R. Apsingekar and P. L. De Leon",
	title="{S}peaker {M}odel {C}lustering for {E}fficient {S}peaker {I}dentification in {L}arge {P}opulation {A}pplications",
	journal="IEEE Trans. on Speech and Language Processing",
	volume="17",
	pages="848-853",
	year="2009"
}

@inproceedings{marc2009,
      author = { M. Ferras and C. Barras and J. L. Gauvain},
      title= {{L}attice-based {MLLR} for {S}peaker {R}ecognition},
      booktitle={Proc. of ICASSP},
      year = {2009},
      pages={4537-4540}
}

@article{ferras_ieee2010,
	Author = {M. Ferras and C.C. Leung and C. Barras and J.-L. Gauvain},
	Journal = {IEEE Trans. on Audio, Speech and Language Processing},
	Pages = {1366-1378},
	Title = {{C}omparison of {S}peaker {A}daptation {M}ethods as {F}eature {E}xtraction for {SVM} Based {S}peaker {R}ecognition},
	Volume = {18},
	Year = {2010}}

@inproceedings{achintya-odyssey2010,
	author= {A. K. Sarkar and S. Umesh},
	title= {{I}nvestigation of {S}peaker-{C}lustered {UBM}s based on {V}ocal {T}ract {L}engths and {MLLR} matrices for {S}peaker {V}erification},
	booktitle={Proc. of Odyssey Speaker and Language Recognition Workshop},
	pages= {2738-2741},
	year = {2010}

}

@inproceedings{achintya-odyssey2010_spkid,
	      author= {A. K. Sarkar and S. Umesh and S. P. Rath},
	      title= {{C}omputationally {E}fficient {S}peaker {I}dentification for {L}arge {P}opulation {T}asks using {MLLR} and {S}ufficient {S}tatistics},
	      booktitle={Proc. of Odyssey Speaker and Language Recognition Workshop},
	      pages= {7-11},
              year = {2010}
}


@inproceedings{Zhang-odyssey2010,
       author= {W. Q. Zhang and Y. Shan and J. Liu},
       title= {{M}ultiple {B}ackground {M}odels for {S}peaker {V}erification},
       booktitle={Proc. of Odyssey Speaker and Language Recognition Workshop},
       pages={47-51},
       year = {2010}
}


@inproceedings{achintya-inter2010,
             author= {A. K. Sarkar and S. Umesh},
             title= {{F}ast {C}omputation of {S}peaker {C}haracterization {V}ector using {MLLR} and {S}ufficient {S}tatistics in {A}nchor {M}odel {F}ramework},
             booktitle = {Proc. of INTERSPEECH},
             pages={2738-2741},
	     year = {2010}
}


@inproceedings{Kockmann2012,
            author= {M. Kockmann and L. Burget and O. Glembek and L. Ferrer and J. Cernocky},
            title= {{P}rosodic {S}peaker {V}erification using {S}ubspace {M}ultinomial {M}odels with {I}ntersession {C}ompensation},
            booktitle = {Proc. of INTERSPEECH},
            pages= {1061-1064},
            year= {2010}
           
}

@inproceedings{achintya-icassp2011,
	        author= {A. K. Sarkar and S. Umesh},
	        title= {{U}se of {VTL}-wise {M}odels in {F}eature-{M}apping {F}ramework to {A}chieve {P}erformance of {M}ultiple-{B}ackground {M}odels in {S}peaker {V}erification},
	        booktitle={Proc. of IEEE Int. Conf. Acoust. Speech Signal Processing (ICASSP)},
	        year = {2011}

}

@article{Deka_ieee2011,
	author = {Dehak, N. and Kenny, P. J and Dehak, R. and Dumouchel, P. and Ouellet, P.},
	Journal = {IEEE Trans. on Audio, Speech and Language Processing},
	Pages = {788--798},
	Title = {{F}ront-{E}nd {F}actor {A}nalysis for {S}peaker {V}erification},
	Volume = {19},
	number = {4},
	Year = {2011}}

@inproceedings{achintya-interspeech2011,
	author= {A. K. Sarkar and S. Umesh},
	title= {{E}igen-voice {B}ased {A}nchor {M}odeling {S}ystem for {S}peaker {I}dentification  using  {MLLR} {S}uper-vector},
        booktitle={Proc. of INTERSPEECH},
        pages={2357-2360},
	year = {2011}
}

@inproceedings{Pierre-interspeech2011,
	author= {P. M. Bousquet and D. Matrouf and J. F. Bonastre},
        title= {{I}ntersession {C}ompensation and {S}coring {M}ethods in the i-vectors {S}pace for {S}peaker {R}ecognition},
	booktitle={Proc. of INTERSPEECH},
        pages={485-488},
	year = {2011}
}       


@inproceedings{SenoussaouiInterspch2011,
	Author = {Senoussaoui, M. and Kenny, P. and BrŸmmer, N. and Villiers, E. de and Dumouchel, P.},
	Booktitle = {Proc. of Interspeech},
	Pages = {25--28},
	Title = {Mixture of {PLDA} Models in I-Vector Space for Gender-Independent Speaker Recognition},
	Year = {2011}}


@inproceedings{Nicolas_Scheffer2011,
	author= {Nicolas Scheffer and Yun Lei and Luciana Ferrer},
	title={{F}actor {A}nalysis {B}ack {E}nds for {MLLR} {T}ransforms in {S}peaker {R}ecognition},
	booktitle={Proc. of INTERSPEECH},
        pages= {257-260},
	year = {2011}
}

@inproceedings{achintya-icassp2012,
                author= {A. K. Sarkar and S. Umesh and J. F. Bonastre},
                title= {{C}omputationally {E}fficient {S}peaker {I}dentification using {F}ast-{MLLR}  Based {A}nchor {M}odeling},
                booktitle={Proc. of IEEE Int. Conf. Acoust. Speech Signal Processing (ICASSP)},
                year = {2012}

}

@inproceedings{achintya-eusipco2012,
     author={A. K. Sarkar and J. F. Bonastre and D. Matrouf},
     title={{S}peaker {V}erification {u}sing {m}-vector {E}xtracted from {MLLR} {S}uper-vector},
     booktitle={Proc. of 20th European Signal Processing Conference (EUSIPCO)},
     pages={21-25},
     year = {2012}
} 

@inproceedings{achintya-interspeech2012,
        author={A. K. Sarkar and D. Matrouf and P. M. Bousquet and J. F. Bonastre},
        title={{S}tudy of the {E}ffect of {I}-vector {M}odeling on {S}hort and {M}ismatch {U}tterance {D}uration for {S}peaker {V}erification},
        booktitle={Proc. of INTERSPEECH},
        pages= {2662-2665},
        year = {2012}
}

@inproceedings{Larcher-icassp2012,
         author={A. Larcher and  P. Bousquet and  K. Lee and  D. Matrouf and  H. Li and  J. F. Bonastre},
         title={{I}-vectors in the {C}ontext of {P}honetically-constrained {S}hort {U}tterances for {S}peaker {V}erification},
         booktitle={Proc. of IEEE Int. Conf. Acoust. Speech Signal Processing (ICASSP)},
         pages = {4773-4776}, 
         year = {2012}
}


@inproceedings{Pierre-odyssey2012,
       author= {P. M. Bousquet and A. Larcher and D. Matrouf and J. F. Bonastre and O. Plchot},
       title= {{V}ariance-{S}pectra {B}ased {N}ormalization for i-vector {S}tandard and {P}robabilistic {L}inear {D}iscriminant {A}nalysis},
       booktitle={Proc. of Odyssey Speaker and Language Recognition Workshop},
       year= {2012}
}



@inproceedings{Prince2012,
        author= {Simon J.D. Prince},
        title= {{C}omputer {V}ision: {M}odels {L}earning and {I}nference},
        booktitle={{Cambridge University Press}},
        year=2012
}

@inproceedings{Bouallegue2012,
        author= {M. Bouallegue and E. Ferreira and D. Matrouf and G. Linares and M. Goudi and P. Nocera},
        title= {{A}coustic {M}odeling for {U}nder-resourced {L}anguages based on {V}ectorial {HMM}-states representation using {S}ubspace {G}aussian {M}ixture {M}odels},
        booktitle={Proc. of Spoken Language Technology Workshop},
        pages={330-335},
        year={2012}
}

@inproceedings{Gregor2012,
       author={ G. Dupuy and M. Rouvier and S. Meignier and Y.  Estève},
       title= {{I}-vectors and {ILP} {C}lustering {A}dapted to {C}ross-show {S}peaker {D}iarization},
       booktitle={Proc. of INTERSPEECH},
       pages={2174-2177},
       year={2012}
}



@inproceedings{achintya-icassp2013,
author={A. K. Sarkar and C. Barras and V. B. Le},
title={{L}attice {MLLR} based \emph{m-vector} {S}ystem for {S}peaker {V}erification},
booktitle={Proc. of IEEE Int. Conf. Acoust. Speech Signal Processing (ICASSP)},
pages= {7654 - 7658},
year = {2013}

}


@inproceedings{Soufifar2013,
author={M. M. Soufifar and  L. Burget and O. Plchot and S. Cumani and J. Cernocky},
title={{R}egularized {S}ubspace n-gram {M}odel for {P}honotactic {I}-vector {E}xtraction},
booktitle={Proc. of INTERSPEECH},
pages={74-78},
year= {2013}
}


@inproceedings{Taufiq-icassp2013,
author={T. Hasan and R. Saeidi and  J. H. L. Hansen and  D. A. van Leeuwen},
title={{Duration Mismatch Compensation for i-vector based Speaker Recognition System}},
booktitle={Proc. of IEEE Int. Conf. Acoust. Speech Signal Processing (ICASSP)},
pages= {7663 - 7667},
year = {2013}

}

@Article{mit_database,
  title="http://groups.csail.mit.edu/sls/mdsvc"
}

@Book{duda_hart_patt,
	  author = {R.O. Duda and P.E. Hart and D.G. Stork},
	  title  = {{P}attern {C}lassification},
	  publisher = {New York: John Wiley \& Sons},
	  year    = {2001}
}


@misc{Focal,
	author= {N. Brummer},
	title= {{F}oCal {T}oolkit},
	 howpublished ={Available at http://sites.google.com/site/nikobrummer/focal}
}

@misc{Lnknet,
title= {http://www.ll.mit.edu/mission/communications/ist/lnknet/index.html}
}


@misc{nisteval08,
	Author = {NIST},
	Howpublished = {\url{http://www.itl.nist.gov/iad/mig/tests/sre/2008/sre08_evalplan_release4.pdf}},
	Title = {{T}he {NIST} {Y}ear 2008 {S}peaker {R}ecognition {E}valuation {P}lan.},
	Year = 2008}

@misc{nisteval10,
	Author = {NIST},
	Howpublished = {\url{http://www.itl.nist.gov/iad/mig//tests/sre/2010/NIST_SRE10_evalplan.r6.pdf}},
	Title = {{T}he {NIST} {Y}ear 2010 {S}peaker {R}ecognition {E}valuation {P}lan.},
	Year = 2010}


@inproceedings{Cumani2012,
	title = {Independent component analysis and {MLLR} transforms for speaker identification},
	url = {http://ieeexplore.ieee.org/xpls/abs_all.jsp?arnumber=6288886},
	urldate = {2014-04-08},
	booktitle = {Acoustics, Speech and Signal Processing ({ICASSP)}, 2012 {IEEE} International Conference on},
	publisher = {{IEEE}},
	author = {Cumani, Sandro and Plchot, Oldrich and Karafi‡t, Martin},
	year = {2012},
	pages = {4365Ð4368}}


@inproceedings{Stolcke2012,
	title = {Speaker recognition with region-constrained {MLLR} transforms},
	url = {http://ieeexplore.ieee.org/xpls/abs_all.jsp?arnumber=6288894},
	urldate = {2014-04-08},
	booktitle = {Acoustics, Speech and Signal Processing ({ICASSP)}, 2012 {IEEE} International Conference on},
	publisher = {{IEEE}},
	author = {Stolcke, Andreas and Mandal, Arindam and Shriberg, Elizabeth},
	year = {2012},
	pages = {4397Ð4400}}


\begin{thebibliography}{10}

\bibitem{reynold00}
D.~A. Reynolds, T.~F. Quatieri, and R.~B. Dunn,
\newblock ``{S}peaker {V}erification using {A}dapted {G}aussian {M}ixture
  {M}odels,''
\newblock {\em Digital Signal Processing}, vol. 10, pp. 19--41, 2000.

\bibitem{Deka_ieee2011}
N.~Dehak, P.~J Kenny, R.~Dehak, P.~Dumouchel, and P.~Ouellet,
\newblock ``{F}ront-{E}nd {F}actor {A}nalysis for {S}peaker {V}erification,''
\newblock {\em IEEE Trans. on Audio, Speech and Language Processing}, vol. 19,
  no. 4, pp. 788--798, 2011.

\bibitem{Stolcke05}
A.~Stolcke, L.~Ferrer, S.~Kajarekar, E.~Shriberg, and A.~Venkataraman,
\newblock ``{MLLR} {T}ransforms as {F}eatures in {S}peaker {R}ecognition,''
\newblock in {\em Proc. of EUROSPEECH}, September 2005, pp. 2425--2428.

\bibitem{KaramIcassp2008}
Z.~N. Karam and W.~M. Campbell,
\newblock ``{A} {M}ulti-class {MLLR} {K}ernel for {SVM} {S}peaker
  {R}ecognition,''
\newblock in {\em Proc. of ICASSP}, 2008, pp. 4117--4120.

\bibitem{ferras_ieee2010}
M.~Ferras, C.C. Leung, C.~Barras, and J.-L. Gauvain,
\newblock ``{C}omparison of {S}peaker {A}daptation {M}ethods as {F}eature
  {E}xtraction for {SVM} based {S}peaker {R}ecognition,''
\newblock {\em IEEE Trans. on Audio, Speech and Language Processing}, vol. 18,
  pp. 1366--1378, 2010.

\bibitem{Nicolas_Scheffer2011}
Nicolas Scheffer, Yun Lei, and Luciana Ferrer,
\newblock ``{F}actor {A}nalysis {B}ack {E}nds for {MLLR} {T}ransforms in
  {S}peaker {R}ecognition,''
\newblock in {\em Proc. of INTERSPEECH}, 2011, pp. 257--260.

\bibitem{Cumani2012}
Sandro Cumani, Oldrich Plchot, and Martin Karafi‡t,
\newblock ``Independent component analysis and {MLLR} transforms for speaker
  identification,''
\newblock in {\em Acoustics, Speech and Signal Processing ({ICASSP)}, 2012
  {IEEE} International Conference on}. 2012, p. 4365Ð4368, {IEEE}.

\bibitem{Stolcke2012}
Andreas Stolcke, Arindam Mandal, and Elizabeth Shriberg,
\newblock ``Speaker recognition with region-constrained {MLLR} transforms,''
\newblock in {\em Acoustics, Speech and Signal Processing ({ICASSP)}, 2012
  {IEEE} International Conference on}. 2012, p. 4397Ð4400, {IEEE}.

\bibitem{achintya-eusipco2012}
A.~K. Sarkar, J.~F. Bonastre, and D.~Matrouf,
\newblock ``{S}peaker {V}erification {u}sing {m}-vector {E}xtracted from {MLLR}
  {S}uper-vector,''
\newblock in {\em Proc. of 20th European Signal Processing Conference
  (EUSIPCO)}, 2012, pp. 21--25.

\bibitem{achintya-icassp2013}
A.~K. Sarkar, C.~Barras, and V.~B. Le,
\newblock ``{L}attice {MLLR} based \emph{m-vector} {S}ystem for {S}peaker
  {V}erification,''
\newblock in {\em Proc. of IEEE Int. Conf. Acoust. Speech Signal Processing
  (ICASSP)}, 2013, pp. 7654 -- 7658.

\bibitem{nisteval08}
NIST,
\newblock ``{T}he {NIST} {Y}ear 2008 {S}peaker {R}ecognition {E}valuation
  {P}lan.,''
  \url{http://www.itl.nist.gov/iad/mig/tests/sre/2008/sre08_evalplan_release4.pdf},
  2008.

\bibitem{nisteval10}
NIST,
\newblock ``{T}he {NIST} {Y}ear 2010 {S}peaker {R}ecognition {E}valuation
  {P}lan.,''
  \url{http://www.itl.nist.gov/iad/mig//tests/sre/2010/NIST_SRE10_evalplan.r6.pdf},
  2010.

\bibitem{Leggeter95}
C.~Leggetter and P.~Woodland,
\newblock ``{M}aximum {L}ikelihood {L}inear {R}egression for {S}peaker
  {A}daptation of {HMM}s,''
\newblock {\em Computer Speech and Language}, vol. 9, pp. 171--186, 1995.

\bibitem{padmanabhan_00}
M.~Padmanabhan, G.~Saon, and G.~Zweig,
\newblock ``{L}attice-{B}ased {U}nsupervised {MLLR} for {S}peaker
  {A}daptation,''
\newblock in {\em Proc. of the ISCA ITRW ASR2000}, 2000, pp. 128--131.

\bibitem{uebel_01}
L.F. Uebel and P.C. Woodland,
\newblock ``{I}mprovements in {L}inear {T}ranformation based {S}peaker
  {A}daptation,''
\newblock in {\em Proc. of IEEE Int. Conf. Acoust. Speech Signal Processing
  (ICASSP)}, 2001, pp. 49--52.

\bibitem{marc2009}
M.~Ferras, C.~Barras, and J.~L. Gauvain,
\newblock ``{L}attice-based {MLLR} for {S}peaker {R}ecognition,''
\newblock in {\em Proc. of ICASSP}, 2009, pp. 4537--4540.

\bibitem{Gregor2012}
G.~Dupuy, M.~Rouvier, S.~Meignier, and Y.~Estève,
\newblock ``{I}-vectors and {ILP} {C}lustering {A}dapted to {C}ross-show
  {S}peaker {D}iarization,''
\newblock in {\em Proc. of INTERSPEECH}, 2012, pp. 2174--2177.

\bibitem{Kockmann2012}
M.~Kockmann, L.~Burget, O.~Glembek, L.~Ferrer, and J.~Cernocky,
\newblock ``{P}rosodic {S}peaker {V}erification using {S}ubspace {M}ultinomial
  {M}odels with {I}ntersession {C}ompensation,''
\newblock in {\em Proc. of INTERSPEECH}, 2010, pp. 1061--1064.

\bibitem{Soufifar2013}
M.~M. Soufifar, L.~Burget, O.~Plchot, S.~Cumani, and J.~Cernocky,
\newblock ``{R}egularized {S}ubspace n-gram {M}odel for {P}honotactic
  {I}-vector {E}xtraction,''
\newblock in {\em Proc. of INTERSPEECH}, 2013, pp. 74--78.

\bibitem{Larcher-icassp2012}
A.~Larcher, P.~Bousquet, K.~Lee, D.~Matrouf, H.~Li, and J.~F. Bonastre,
\newblock ``{I}-vectors in the {C}ontext of {P}honetically-constrained {S}hort
  {U}tterances for {S}peaker {V}erification,''
\newblock in {\em Proc. of IEEE Int. Conf. Acoust. Speech Signal Processing
  (ICASSP)}, 2012, pp. 4773--4776.

\bibitem{Taufiq-icassp2013}
T.~Hasan, R.~Saeidi, J.~H.~L. Hansen, and D.~A. van Leeuwen,
\newblock ``{Duration Mismatch Compensation for i-vector based Speaker
  Recognition System},''
\newblock in {\em Proc. of IEEE Int. Conf. Acoust. Speech Signal Processing
  (ICASSP)}, 2013, pp. 7663 -- 7667.

\bibitem{Pierre-interspeech2011}
P.~M. Bousquet, D.~Matrouf, and J.~F. Bonastre,
\newblock ``{I}ntersession {C}ompensation and {S}coring {M}ethods in the
  i-vectors {S}pace for {S}peaker {R}ecognition,''
\newblock in {\em Proc. of INTERSPEECH}, 2011, pp. 485--488.

\bibitem{Prince2012}
Simon~J.D. Prince,
\newblock ``{C}omputer {V}ision: {M}odels {L}earning and {I}nference,''
\newblock in {\em {Cambridge University Press}}, 2012.

\bibitem{SenoussaouiInterspch2011}
M.~Senoussaoui, P.~Kenny, N.~BrŸmmer, E.~de Villiers, and P.~Dumouchel,
\newblock ``Mixture of {PLDA} models in i-vector space for gender-independent
  speaker recognition,''
\newblock in {\em Proc. of Interspeech}, 2011, pp. 25--28.

\bibitem{Pierre-odyssey2012}
P.~M. Bousquet, A.~Larcher, D.~Matrouf, J.~F. Bonastre, and O.~Plchot,
\newblock ``{V}ariance-{S}pectra {B}ased {N}ormalization for i-vector
  {S}tandard and {P}robabilistic {L}inear {D}iscriminant {A}nalysis,''
\newblock in {\em Proc. of Odyssey Speaker and Language Recognition Workshop},
  2012.

\bibitem{Prasad05}
R.~Prasad et~al.,
\newblock ``{T}he 2004 {BBN}/{LIMSI} {20xRT} {E}nglish {C}onversational
  {T}elephone {S}peech {R}ecognition {S}ystem,''
\newblock in {\em Proc. of INTERSPEECH}, 2005, pp. 1645--1648.

\bibitem{Fousek2008}
P.~Fousek, L.~Lamel, and J.~L. Gauvain,
\newblock ``{T}ranscribing {B}roadcast {D}ata using {MLP} {F}eatures,''
\newblock in {\em Proc. of INTERSPEECH}, 2008, pp. 1433--1436.

\end{thebibliography}


@Article{Lamp86,
  author =       "A.B. Smith and C.D. Jones and E.F. Roberts",
  title =        "Article Title",
  journal = 	 "Journal",
  year = 	 "1920",
  volume = 	 "62",
  pages = 	 "291-294",
  month = 	 "January"
}
\end{document}